\documentclass[letterpaper,twocolumn,aps,prl,superscriptaddress,amsmath,amssymb,floatfix]{revtex4-2}
\usepackage{mathptmx}
\usepackage[latin9]{inputenc}
\usepackage{color}
\usepackage{amsmath}
\usepackage{amssymb}
\usepackage{graphicx}
\usepackage{esint}
\usepackage{comment}
\usepackage[unicode=true,
 bookmarks=true,bookmarksnumbered=false,bookmarksopen=false,
 breaklinks=false,pdfborder={0 0 1},backref=false,colorlinks=true]
 {hyperref}
\hypersetup{
 linkcolor=magenta,urlcolor=blue,citecolor=blue,pdfstartview={FitH},hyperfootnotes=false}

\makeatletter

\usepackage{textcomp}
\usepackage{epstopdf}

\pdfpageheight\paperheight
\pdfpagewidth\paperwidth

\@ifundefined{textcolor}{}{%
 \definecolor{BLACK}{gray}{0}
 \definecolor{WHITE}{gray}{1}
 \definecolor{RED}{rgb}{1,0,0}
 \definecolor{GREEN}{rgb}{0,1,0}
 \definecolor{BLUE}{rgb}{0,0,1}
 \definecolor{CYAN}{cmyk}{1,0,0,0}
 \definecolor{MAGENTA}{cmyk}{0,1,0,0}
 \definecolor{YELLOW}{cmyk}{0,0,1,0}
}

\usepackage{xcolor}
\usepackage{soul}
\usepackage{lineno}
\modulolinenumbers[5] 

\newcommand{\bra}[1]{\ensuremath{\left\langle#1\right|}}
\newcommand{\ket}[1]{\ensuremath{\left|#1\right\rangle}}

\definecolor{blue}{rgb}{0,0,1}
\definecolor{red}{rgb}{1,0,0}
\definecolor{green}{rgb}{0,1,0}

\usepackage{soul}

\makeatother
\begin{document}
	
\title{Extending coherence time beyond break-even point using only drives and dissipation}

\author{Lida Sun}
\thanks{These authors contributed equally to this work.}
\author{Yifang Xu}
\thanks{These authors contributed equally to this work.}
\affiliation{Center for Quantum Information, Institute for Interdisciplinary Information Sciences, Tsinghua University, Beijing 100084, China}

\author{Yilong Zhou}
\thanks{These authors contributed equally to this work.}
\affiliation{Center for Quantum Information, Institute for Interdisciplinary Information Sciences, Tsinghua University, Beijing 100084, China}

\author{Ziyue Hua}
\affiliation{Center for Quantum Information, Institute for Interdisciplinary Information Sciences, Tsinghua University, Beijing 100084, China}

\author{Weiting Wang}
\affiliation{Center for Quantum Information, Institute for Interdisciplinary Information Sciences, Tsinghua University, Beijing 100084, China}

\author{Jie Zhou}
\affiliation{Center for Quantum Information, Institute for Interdisciplinary Information Sciences, Tsinghua University, Beijing 100084, China}

\author{Zi-Jie Chen}
\affiliation{Laboratory of Quantum Information, University of Science and Technology of China, Hefei 230026, China}

\author{Lui Zuccherelli de Paula}
\affiliation{Laboratory of Quantum Information, University of Science and Technology of China, Hefei 230026, China}

\author{Qing-Xuan Jie}
\affiliation{Laboratory of Quantum Information, University of Science and Technology of China, Hefei 230026, China}

\author{Guangming~Xue}
\affiliation{Beijing Academy of Quantum Information Sciences, Beijing 100084, China}
\affiliation{Hefei National Laboratory, Hefei 230088, China}

\author{Haifeng~Yu}
\email{hfyu@baqis.ac.cn}
\affiliation{Beijing Academy of Quantum Information Sciences, Beijing 100084, China}
\affiliation{Hefei National Laboratory, Hefei 230088, China}

\author{Weizhou Cai}
\email{caiwz@ustc.edu.cn}
\affiliation{Laboratory of Quantum Information, University of Science and Technology of China, Hefei 230026, China}

\author{Chang-Ling Zou}
\email{clzou321@ustc.edu.cn}
\affiliation{Laboratory of Quantum Information, University of Science and Technology of China, Hefei 230026, China}
\affiliation{Anhui Province Key Laboratory of Quantum Network, University of Science and Technology of China, Hefei 230026, China}
\affiliation{Hefei National Laboratory, Hefei 230088, China}
\author{Luyan Sun}
\email{luyansun@tsinghua.edu.cn}
\affiliation{Center for Quantum Information, Institute for Interdisciplinary Information Sciences, Tsinghua University, Beijing 100084, China}
\affiliation{Hefei National Laboratory, Hefei 230088, China}

\begin{abstract}

\textbf{Quantum error correction (QEC) aims to mitigate the loss of quantum information to the environment, which is a critical requirement for practical quantum computing. 
Existing QEC implementations heavily rely on measurement-based feedback, however, constraints on readout fidelity, hardware latency, and system complexity often limit both performance and scalability. 
Autonomous QEC (AQEC) seeks to overcome these obstacles by stabilizing logical codewords using introduced drives that provide coherent control and engineered dissipation.
Here, we propose an AQEC protocol, derived from quantum channel simulation, that is applicable to arbitrary error-correcting codes.
As a demonstration, we implement the protocol using a binomial code encoded in a long-lived bosonic mode (lifetime $>$ 1~ms), and extend the logical qubit coherence time to 1.04 times that of the best physical qubit in the system.
This is the first experimental realization of an AQEC-protected bosonic logical qubit beyond the break-even point, proving that coherence time can indeed be extended by introducing only drives and dissipation.
Our results highlight the performance and scalability potential of AQEC, marking an important step toward large-scale, universal quantum computing.}

\end{abstract}

\maketitle
\noindent \textbf{\large{}{}Introduction}{\large\par}

\noindent The inevitable decoherence of quantum information poses a central obstacle to practical quantum computing~\cite{QC_Nielsen_2010,Preskill2025}. Traditional strategies to combat this fundamental challenge have pursued two complementary approaches: optimizing materials and shields to isolate qubits from environment noise, or developing physical qubits with intrinsically longer coherence times, such as nuclear spins~\cite{Pla2013,BradleyPhysRevX2019,WangPRXQuantum2025}. Alternatively, the principle of quantum information offers an elegant solution to this problem through quantum error correction (QEC)~\cite{QEC_Shor_1995, QEC_Steane_1996, QEC_Fowler_2012}, which exploits enlarged Hilbert spaces to encode and protect logical qubits via sophisticated quantum control techniques. Active QEC relies on repeated syndrome measurements followed by real-time feedback operations, achieving remarkable recent successes including the break-even milestone~\cite{CatBreakeven_Ofek_2016,BinomialBreakeven_Ni_2023,GKPBreakeven_Sivak_2023,Acharya2025Nature}. However, this measurement-based approach requires demanding hardware and suffers from feedback latencies and measurement errors. In contrast, the passive approach, known as autonomous QEC (AQEC)~\cite{Kerckhoff2010,Leghtas_2013,Kapit2016,Lihm2018,Shtanko2025,Kruckenhauser2025}, promises to overcome these limitations by engineering dissipation that generates spontaneous emission-like transitions to recover quantum information back to the code subspace, without requiring measurements on the system. Despite many efforts in the past few years, a fundamental question remains open~\cite{ZhaoyouWang2022PRXQ}: Can we extend the coherence time of a quantum system by using only drives and dissipation?


Mathematically, decoherence due to continuous interaction with environments can be described by the master equation in Lindblad form~\cite{Lindblad1976}:
\begin{equation}
    \frac{d\rho}{dt}=\sum_j\gamma_j\mathcal{L}[E_j]\rho,
\end{equation}
where $\mathcal{L}[E_j]\rho=E_j\rho E_j^\dagger-\frac{1}{2}\{E_j^\dagger E_j,\rho\}$ is the Lindblad superoperator describing how the error operator $E_j$ corrupts the quantum state $\rho$ at rate $\gamma_j$. For a QEC code with code subspace projector $\mathcal{P}_C$ and error subspace projectors $\mathcal{P}_j$, the Knill-Laflamme condition~\cite{KL1997} $\mathcal{P}_C E_j^\dagger E_k \mathcal{P}_C = \delta_{jk} \alpha_j \mathcal{P}_C$ ensures that distinct errors map the code subspace to orthogonal error subspaces. 
Continuous recovery can then be implemented by introducing dissipators of the form $\sum_j\Gamma_j\mathcal{L}[R_j]\rho$ into the master equation~\cite{Rojkov2025}, where each recovery dissipator $R_j$ acts at rate $\Gamma_j$ such that $R_j E_j \propto \mathcal{P}_C$ (returning errors to the code subspace) and $\mathcal{P}_C R_j R_k^\dagger \mathcal{P}_C = \delta_{jk} \beta_j \mathcal{P}_C$ (ensuring orthogonality among recovery operations).
Then, the effective decoherence rate is suppressed to $\gamma_{\text{eff},j} = {\gamma_j^2}/({\gamma_j + \Gamma_j})$, indicating passive protection of quantum information by the AQEC process. Nevertheless, practical implementation of AQEC remains highly challenging because it demands precise synthesis of complex many-body Hamiltonians, requires exotic multi-qubit interactions or high-order multi-photon processes, and necessitates fine-tuning of coupling parameters specifically tailored to a given QEC code~\cite{AQEC_Gertler_2021,AQEC_Li2025,DeBry2025}.

\begin{figure*}[t]
    \centering
    \includegraphics[width=480pt]{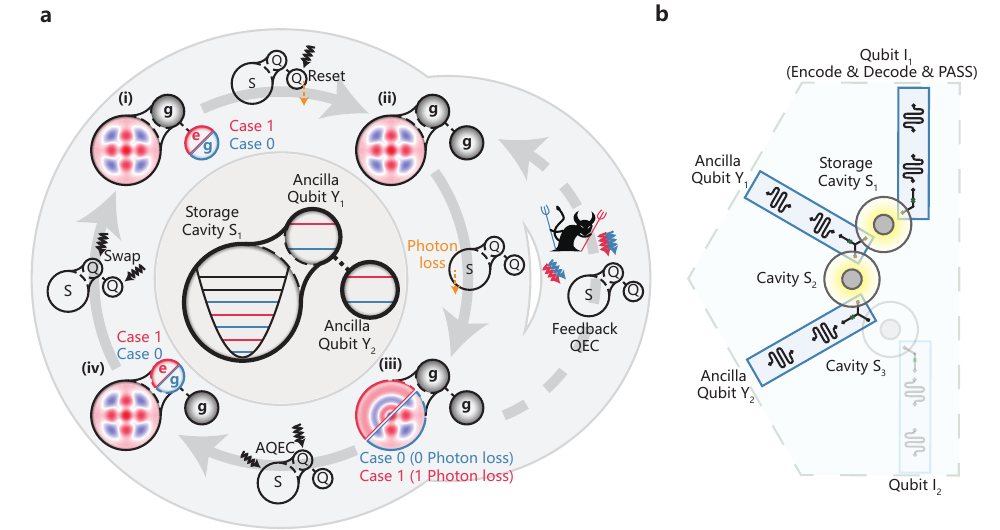}
    \caption{\textbf{Principle and experimental device for the AQEC protocol.} \textbf{a.} Conceptual illustration of the AQEC protocol. Our protocol incorporates a storage cavity $S_1$ and two ancilla qubits $Y_1$ and $Y_2$, as shown in the central panel. Each AQEC cycle proceeds clockwise from the initial state (i). First, ancilla qubit $Y_2$ is reset to its ground state to remove error entropy from the previous cycle, purifying the whole composite system (ii). The logical state then undergoes a 220~$\mu$s free evolution period, where it probabilistically accumulates single-photon-loss errors, projecting it into the error subspace (iii). An AQEC recovery gate is then applied to coherently map these errors on the logical qubit onto an excitation of ancilla qubit $Y_1$ (iv). Finally, the entropy in $Y_1$ is transferred to the distal qubit $Y_2$ via a coherent Swap gate, preventing unwanted back-action on the logical state in $S_1$. For comparison, the right branch illustrates a conventional measurement-feedback QEC protocol. \textbf{b.} Schematic diagram of our experimental device. The concentric circles represent the 3D cavities, and the blue rectangles represent chips housing transmon qubits. Qubit $I_1$ is used for encoding and decoding logical states and for applying the PASS drives. Cavity $S_2$ operates as a coupler between the two ancilla qubits $Y_1$ and $Y_2$, enabling a high-fidelity swap of excitations. 
    }
    \label{fig:concept}
\end{figure*}

\begin{figure*}[t]
    \centering
    \includegraphics[width=480pt]{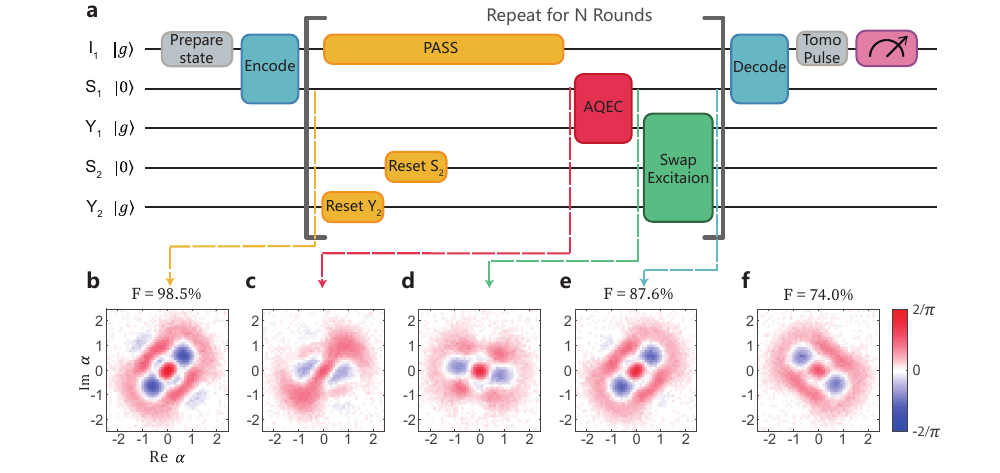}
    \caption{\textbf{The AQEC process.}
    \textbf{a.} Experimental sequences for the AQEC protocol. 
    \textbf{b.} Wigner tomography of the logical state $\ket{-Y_L}$ in cavity $S_1$ after encoding (fidelity = 98.5\%).
    \textbf{c.} Wigner tomography of $\ket{-Y_L}$ after 220~$\mu$s of free evolution with PASS drives.
    \textbf{d.} Wigner tomography of $\ket{-Y_L}$ before the Swap pulse in the first AQEC round.
    \textbf{e.} Wigner tomography of $\ket{-Y_L}$ after one full round of AQEC (fidelity = 87.6\%).
    \textbf{f.} For comparisons, Wigner tomography of $\ket{-Y_L}$ after 220~$\mu$s of free evolution without PASS drives, showing significant decay (fidelity = 74.0\%). 
    }
    \label{fig:sequence}
\end{figure*}

Here, we answer the key question with a definitive ``yes". We propose and demonstrate a new approach to realize AQEC by simulating arbitrary dissipation processes via quantum channel simulation~\cite{ShenPRB2017,Hu2018channel,CaiAQuO}. This method provides a scalable pathway for QEC that requires neither measurement-based feedback nor precise Hamiltonian engineering. By introducing an ancilla qubit and implementing a Stinespring dilation unitary~\cite{QC_Nielsen_2010}, we realize effective continuous recovery dissipators through carefully designed hierarchical entropy management. Employing a high-dimensional bosonic mode to encode quantum information with optimized control pulse sequences, we successfully extend the coherence time of a logical qubit to beyond the break-even point of QEC. Our AQEC approach is applicable to arbitrary recovery dissipators and can be generalized to other QEC codes with more recovery dissipators and larger dimensions, thereby providing a pathway to protect quantum information without exotic Hamiltonian synthesis or parameter fine-tuning. 

\smallskip{}

\noindent \textbf{\large{}{}Results}{\large\par}

\noindent 
Instead of engineering the Markovian dissipator directly, we implement repetitive channel simulation using the Trotter approximation~\cite{Han2021}:
\begin{equation}
e^{t\Gamma_j\mathcal{L}[R_j]} \approx \left(e^{\Delta t\Gamma_j\mathcal{L}[R_j]}\right)^{n} \approx \left(\mathcal{R}_j\right)^{n},
\end{equation}
where the recovery channel $\mathcal{R}_j$ is a completely positive trace-preserving (CPTP) map $\mathcal{R}_j(\rho) = R_j\rho R_j^\dagger + R_j^{\prime}\rho R_{j}^{\prime\dagger}$, with $R_j^\dagger R_j+R_{j}^{\prime\dagger}R_j^{\prime}=\mathcal{P}_C+\mathcal{P}_j$. An arbitrary CPTP map can be realized by introducing an extra ancilla (spanning the subspace $\{ \ket{g},\ket{e} \}$) and applying a Stinespring dilation unitary followed by a partial trace over the ancilla, as given by~\cite{QC_Nielsen_2010}:
\begin{equation}
    \mathcal{R}_j(\rho)= \text{Tr}_A[U_j(\ket{g}\bra{g}\otimes\rho)U_j^{\dagger}],
    \label{eq:UAQEC}
\end{equation}
where $\text{Tr}_A$ represents the partial trace over the ancilla and $U_j$ is the dilation unitary applied to the composite system. The effective dissipator for AQEC can be realized by implementing $U_j$ and resetting the ancilla autonomously and repetitively, with $\tau=1/\Gamma_j$ being the repetition period. Considering potential imperfections in realizing the recovery channel (infidelity $\epsilon$), an effective decay rate of the quantum information encoded in the code subspace is suppressed to
\begin{equation}
{\gamma_{\text{eff}} \approx \frac{\epsilon}{\tau} + \frac{\gamma_c\gamma_e\tau}{2}},
\label{eq:gammaeff}
\end{equation}
with $\gamma_{c(e)}$ being the decay rate of states in the code (error) subspace (see Methods for more details). Therefore, $\tau$ should be optimized to balance the occurrence of uncorrectable high-order errors ($\mathcal{O} (\tau)$) and the extra decoherence introduced by the channel with a rate $\propto 1/\tau$. When high-fidelity quantum control is available (i.e., small $\epsilon$), the decoherence rate of the logical state can be suppressed significantly for a small $\tau$. Moreover, the implementation of a given recovery channel is expected to be robust against uncertainties in system parameters, provided that sufficient controllability over the composite system is maintained. Other experimental imperfections primarily contribute to the infidelity $\epsilon$, leading to only mild degradation in the dissipator performance. 


To demonstrate the advantage of our approach in passively protecting complex quantum states, we implement the AQEC protocol on a bosonic binomial code~\cite{BinomialTheory_Micheal_2016}.  Figure~\ref{fig:concept}a illustrates the principle of the AQEC protocol, 
with the corresponding experimental system shown in Fig.~\ref{fig:concept}b (see Methods for more details). Our experimental device consists of three 3D microwave cavities~\cite{Axline2016,Reagor2016,10msCav_Milul_2023,Kim2025} and four transmon qubits~\cite{Transmon_Koch_2007}, featuring three key components: a high-coherence bosonic storage cavity ($S_1$) that hosts the logical qubit, a primary ancilla qubit ($Y_1$) that directly couples to the storage cavity to coherently extract error entropy, and a secondary ancilla qubit ($Y_2$) that enables autonomous entropy removal without introducing back-action on the logical state. From a thermodynamics perspective, AQEC can be treated as a cooling process: error entropy is coherently transferred from the system to the ancilla, and then dissipated to a controlled environment by resetting the ancilla. 
While intrinsic ancilla dissipation could in principle remove entropy, it may also introduce imperfections during the execution of the dilation unitary $U_j$, which can degrade quantum information. To mitigate this, we introduce a hierarchical cooling process to autonomously reset the ancilla, enabling continuous entropy removal from the system while preserving quantum coherence. 

For bosonic systems, the dominant error mechanism is single-photon loss. The logical codewords of the binomial code are defined in the Fock state basis as
$\ket{0_\mathrm{L}} = (\ket{0}+\ket{4})/\sqrt{2}$ and $\ket{1_\mathrm{L}} = \ket{2}$. These states are specifically designed so that a single-photon-loss error maps them into orthogonal error states $\ket{0_\mathrm{E}} = \ket{3}$ and $\ket{1_\mathrm{E}} = \ket{1}$, respectively. By leveraging the dispersive coupling between storage cavity $S_1$ and ancilla qubit $Y_1$, high-fidelity universal control over the large Hilbert space of the bosonic mode is allowed with optimal control via the GRAPE algorithm~\cite{GRAPE_Khaneja_2005,Heeres2017,GRAPE_Chen_2025}. This enables the realization of arbitrary $U_j$ on the composite cavity-ancilla system. In particular, we implement control pulses to realize the AQEC gate~\cite{PASS_Ma_2020} that acts as follows: $\ket{ g}\otimes\ket{\psi_\mathrm{L}} \rightarrow \ket{g}\otimes\ket{\psi_\mathrm{L}}$ and $\ket{ g}\otimes\ket{\psi_\mathrm{E}} \rightarrow \ket{e}\otimes\ket{\psi_\mathrm{L}}$, where $\ket{\psi_\mathrm{L}}$ ($\ket{\psi_\mathrm{E}}$) is an arbitrary logical (error) state within the span of $\{\ket{0_\mathrm{L}}$,$\ket{1_\mathrm{L}}\}$ ($\{\ket{0_\mathrm{E}}$,$\ket{1_\mathrm{E}}\}$).

\begin{figure}[t]
    \centering
    \includegraphics[width=250pt]{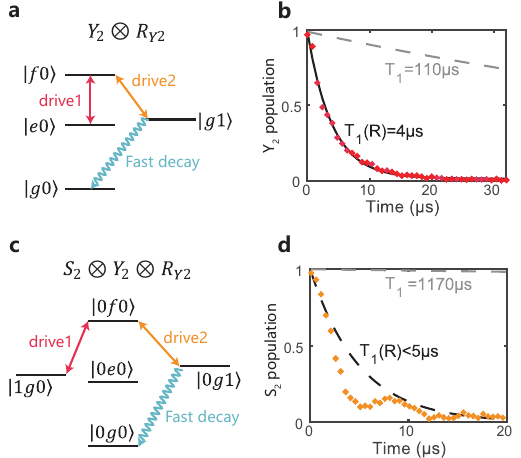}
    \caption{\textbf{Fast reset of the qubit and cavity using a Raman process.}
    \textbf{a.} Principle of the qubit reset scheme. 
    \textbf{b.} Time evolution of qubit $Y_2$ population with the reset drives applied. 
    \textbf{c.} Principle of the cavity reset scheme. 
    \textbf{d.} Time evolution of cavity $S_2$ population with the reset drives applied. 
    }
    \label{fig:reset}
\end{figure}

The experimental sequence for extending quantum coherence and managing entropy is schematically illustrated in Fig.~\ref{fig:concept}a and further detailed in Fig.~\ref{fig:sequence}. A GRAPE-optimized operation is applied to the composite $S_1$-$I_1$ system to encode and decode logical states in $S_1$. Each AQEC cycle begins with the system in a superposition state within the code subspace, while the ancilla qubit $Y_2$ and the bus cavity $S_2$ may have initial excitations [Fig.~\ref{fig:concept}a.(i)]. External drives are then applied to $Y_2$ and $S_2$ to reset both to their ground state [Fig.~\ref{fig:concept}a.(ii), see more details in Fig.~\ref{fig:reset}]. During each cycle, the logical state $\ket{\psi_\mathrm{L}}$ undergoes a free evolution period of $\tau=220\,\mathrm{\mu s}$, which is optimized according to our experimental system (Fig.~S11 in Supplementary Information). Over this interval, the storage cavity $S_1$ probabilistically accumulates single-photon-loss errors, projecting the state into the error subspace $\ket{\psi_\mathrm{E}}$ [Fig.~\ref{fig:concept}a.(iii)]. The free evolution also introduces additional dephasing due to the stochastic nature of single-photon-loss events. To mitigate this effect, we employ a photon-number-resolved a.c. Stark shift (PASS) technique~\cite{PASS_Ma_2020}, which renders the error transparent to the local evolution Hamiltonian. This is achieved by applying a continuous drive to transmon qubit $I_1$ [Fig.~\ref{fig:concept}b], thereby eliminating the extra dephasing (Supplementary Information Sec.~IV for details).  

Following the free evolution period, the AQEC gate is applied to coherently map single-photon-loss errors from the logical qubit onto an excitation of the ancilla qubit $Y_1$ [Fig.~\ref{fig:concept}a.(iv)], thereby effectively extracting error entropy without measuring or disturbing the encoded quantum information. The entropy in $Y_1$ is then transferred to a distal qubit $Y_2$ via a coherent Swap gate, preventing unwanted back-action on $S_1$ that could otherwise arise due to their dispersive coupling. {This hierarchical approach to managing entropy is crucial for achieving break-even in our experiment (see Supplementary Information Sec.~VII)}. {Upon completion of this process, the composite system ($S_1$-$Y_1$) is returned to a purified state, ready for the next round of AQEC.} 
Unlike conventional QEC, which acts as a Maxwell's demon by using measurement and feedback to cool the logical state (Fig.~\ref{fig:concept}a), our repetitive AQEC protocol uses only drives and dissipation.





To experimentally validate our hierarchical entropy management protocol and directly observe the AQEC in action, we employ Wigner functions to visualize the evolution of the logical state, as shown in Figs.~\ref{fig:sequence}b-f. System initialization, purification (see Supplementary Information Sec.~III.B), and quantum state tomography of the storage cavity $S_1$ are implemented via qubit $I_1$ using optimal control pulses. Figure~\ref{fig:sequence}b shows the measured Wigner function of the logical state $\ket{-Y_\mathrm{L}} = (\ket{0_\mathrm{L}}-\mathrm{i}\ket{1_\mathrm{L}})/\sqrt{2}$ immediately after encoding, exhibiting the characteristic interference fringes. The state fidelity of 98.5\%, obtained via maximum likelihood estimation, confirms high-quality state preparation. During the subsequent 220~$\mu$s free evolution period, the logical qubit experiences probabilistic single-photon-loss events, resulting in distortion of the Wigner function and degraded interference visibility in Fig.~\ref{fig:sequence}c.  

Figures~\ref{fig:sequence}d and \ref{fig:sequence}e reveal the Wigner functions after the AQEC recovery gate and the Swap gate, respectively. The difference between these two states originates from the phase shifts imparted by the Swap pulse across the Fock states of cavity $S_1$ (Supplementary Information Sec.~V.D). Remarkably, after one complete AQEC cycle, the logical state $\ket{-Y_\mathrm{L}}$ recovers to a fidelity of 87.6\% (Fig.~\ref{fig:sequence}e), substantially higher than the 74.0\% fidelity observed after the same 220~$\mu$s evolution without error transparency (PASS drive off) and AQEC protection (Fig.~\ref{fig:sequence}f). This significant fidelity improvement in a single cycle provides direct experimental confirmation that a fully passive approach can effectively reverse the natural decoherence process.

\begin{figure*}
    \centering
    \includegraphics[width=440pt]{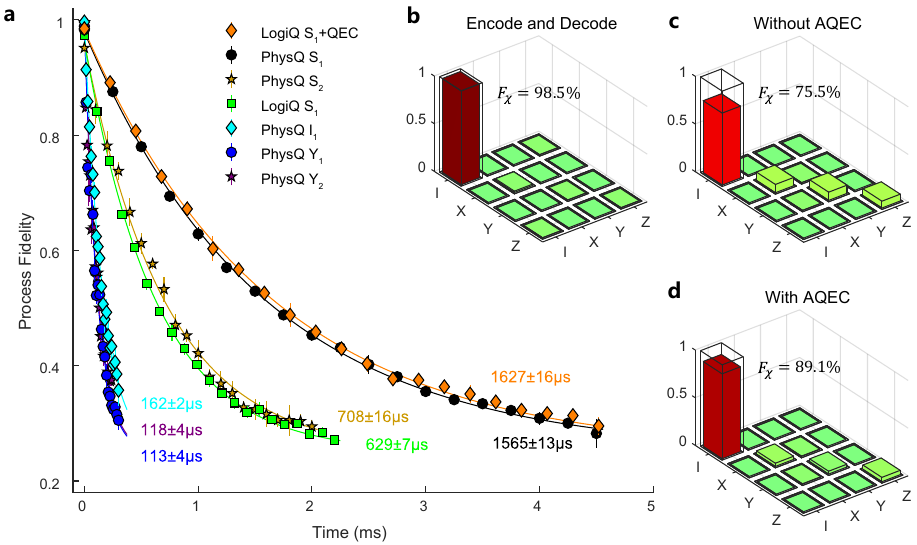}
    \caption{\textbf{AQEC performance.}
    \textbf{a.} Process fidelity over time for different encodings and protocols. The fitted process $T_1$ values are labeled near the curves. LogiQ $S_1$+QEC: binomial encoding of $S_1$ with AQEC; LogiQ $S_1$: standard binomial encoding (no AQEC); PhysQ $S_1$ (or $S_2$, $I_1$, $Y_1$, $Y_2$) : a simple two-level qubit encoded in the lowest two levels of the respective mode. The error bars correspond to standard deviations from multiple measurements. \textbf{b.} Real part of the process matrix for an immediate encode-decode process of the binomial code. \textbf{c.} Real part of the process matrix for the binomial encoding after a $220\,\mathrm{\mu s}$ free evolution without AQEC. \textbf{d.} Real part of the process matrix after one full AQEC cycle.}
    \label{fig:fidelity}
\end{figure*}

In the repetitive AQEC process, one key challenge is to autonomously purge the accumulated entropy in the ancilla qubit while the logical qubit remains protected. As shown in Fig.~\ref{fig:reset}, our reset protocol exploits the hierarchical dissipation rates designed into our system. Following the approach of Ref.~\cite{ETHReset_Magnard_2018}, the effective dissipators can be realized using two-tone driven stimulated Raman-type transitions, where the readout resonator $R_{Y2}$ provides a low-entropy environment. For qubit $Y_2$, which accumulates error entropy during each AQEC cycle, its excitation can be dumped through the dissipator $\mathcal{L}[\ket{g}\bra{e}]\rho$
at a rate $\Omega^2/\gamma_\mathrm{R}$. Here, $\Omega$ is the coupling strength of the Raman transition $\ket{\mathrm{e,0}}\leftrightarrow\ket{\mathrm{g,1}}$ and $\gamma_\mathrm{R}=1/67\,\mathrm{ns}$ is the decay rate of $R_{Y2}$. As shown in Fig.~\ref{fig:reset}b, the effective lifetime of $Y_2$ is shortened to $4\,\mathrm{\mu}$s, compared to $110\,\mathrm{\mu}$s without reset drives. 

Cavity $S_2$ requires similar reset due to imperfections in the Swap operation that can leave residual excitations. By operating in the joint Hilbert space of $S_2\otimes Y_2\otimes R_{Y2}$, we engineer a Raman transition pathway $\ket{\mathrm{1,g,0}}\leftrightarrow\ket{\mathrm{0,g,1}}$, as depicted in Fig.~\ref{fig:reset}c. This reduces the single-photon lifetime of $S_2$ from 1170~$\mu$s to below 5~$\mu$s (Fig.~\ref{fig:reset}d), ensuring complete reset within the operational time window. The non-monotonic decay arises from interference of other transition pathways, which does not affect the eventual purging of $S_2$. The remarkable efficiency of these reset protocols, i.e. over 25-fold and 200-fold acceleration in the decay rates, validates the effectiveness of our hierarchical entropy management. 

Finally, we evaluate the AQEC performance using quantum process tomography~\cite{QC_Nielsen_2010}, by preparing qubit $I_1$ in a set of four input states {$\ket{g}$, $\ket{e}$, $(\ket{g}+\ket{e})/\sqrt{2}$, $(\ket{g}-\mathrm{i}\ket{e})/\sqrt{2}$} and compiling the tomography results from all four initial states. Figure~\ref{fig:fidelity}a shows the the temporal evolution of process fidelity $F_\chi$. The first data point corresponds to the encode-decode operations without evolution, yielding $F_\chi = 98.5\%$ for the logical qubit (Fig.~\ref{fig:fidelity}b). After a single 220~$\mu$s free evolution period, the unprotected logical qubit degrades to $F_\chi = 75.5\%$ (Fig.~\ref{fig:fidelity}c). In contrast, under the AQEC protection, the fidelity rebounds to $F_\chi = 89.1\%$ after the same duration (Fig.~\ref{fig:fidelity}d), consistent with the state fidelity results in Fig.~\ref{fig:sequence}. From the second round onward, each AQEC cycle includes reset operations, yet the fidelity decay remains smooth. Fitting the decay by $F_\chi = F_0e^{-t/\tau}+0.25$ over 20 rounds gives a coherence time of $T_{\mathrm{AQEC}}=1627\,\mathrm{\mu s}$ with AQEC, showing a 2.6-fold extension compared to the unprotected coherence time of $629\,\mathrm{\mu s}$. Here, the offset 0.25 accounts for convergence to the maximally mixed state. 

The ultimate advantage of AQEC should be evaluated by the criteria of break-even: whether the logical qubit coherence time exceeds all physical qubits in the system. To establish this comparison rigorously, we characterize the coherence times of all physical encodings in our device by fitting their process fidelity decay.  Among all measured cavities ($S_1$ and $S_2$) and transmons ($I_1$, $Y_1$, and $Y_2$), the longest-lived physical qubit is encoded in the two lowest Fock states $\{\ket{0}, \ket{1}\}$ of cavity $S_1$, with $T_{\mathrm{best}}^{\mathrm{phys}} = 1565\,\mu\mathrm{s}$. Our AQEC-protected logical qubit achieves $T_{\mathrm{AQEC}} = 1627\,\mu\mathrm{s}$, definitively surpassing the break-even point with a gain factor of 1.04.  

\begin{table}
\begin{tabular}{c|c|c} \hline \hline
\textbf{Error Source} & \textbf{no-photon loss} & \textbf{single-photon loss}\\ \hline
Population of the case & 0.773 & 0.220 \\
Intrinsic error & 5.6\% & 3.9\%\\
Recovery error & 4.0\% & 5.7\%\\
Excitation transfer error & 0.8\% & 2.8\%\\
Thermal \& Reset error & 2.0\% & 2.6\%\\ \hline
Total fidelity & \multicolumn{2}{c}{\centering 87.2\%}\\
Measured fidelity & \multicolumn{2}{c}{\centering 87.0\%}\\ \hline
\end{tabular}
\caption{\textbf{Error budget analysis for a single round of AQEC.}
\label{tab:ErrorBudget}
}
\end{table}

To understand the performance limitations and identify paths for improvement, we perform a comprehensive error budget analysis combining numerical simulations with individual calibration measurements (Supplementary Information Sec.~VIII). We categorize errors into four primary channels: (i) intrinsic errors mainly from multi-photon loss and code space orthogonality degradation caused by no-jump errors (5.6\% in no-loss case vs 3.9\% with single-photon loss); (ii) recovery errors from ancilla decoherence during the AQEC gate (4.0\% vs 5.7\%); (iii) excitation transfer errors during the Swap operation (0.8\% vs 2.8\%); and (iv) thermal and reset errors of the ancilla qubits (2.0\% vs 2.6\%). The total predicted single-round fidelity of 87.2\% closely matches our measured 87.0\%, validating our error model and indicating that ancilla coherence represents the dominant limitation. 

Through a simplified rate equation framework (see Methods), the measured single-round fidelity combined with our cavity decay rates predicts an optimal correction period of $\tau=269\,\mathrm{\mu s}$, agreeing with the more stringent results obtained from our simulations (Fig.~S11 in Supplementary Information). This optimal evolution time emerges from a fundamental trade-off between second-order errors and operation-induced errors, revealing a key feature of the discretized approximation to the ideal continuous dissipator. Further improvement appeals to advanced control to suppress the ancilla decoherence, control imperfections, and higher-order Trotter errors during the implementation of unitary $U_j$. The accumulated errors can then be corrected by the AQEC recovery gate.



\noindent \textbf{\large{}{}Discussion}{\large\par}

\noindent 
In conclusion, we report the first experimental demonstration of a bosonic logical qubit protected by autonomous quantum error correction (AQEC) that surpasses the break-even point using only drives and dissipation. Our AQEC protocol is inspired by textbook quantum channel simulation and is, in principle, applicable to arbitrary QEC codes. Leveraging a dispersively coupled ancilla qubit, we achieve universal control over the oscillator-qubit composite system and realize AQEC for a binomial code. To mitigate dephasing errors of the logical qubit induced by dispersive coupling during ancilla reset, we hierarchically transfer the error entropy to a distal ancilla, cleanly decoupling the reset process form the logical qubit.
This separation of correction and reset processes is a key advantage of our AQEC protocol, offering enhanced flexibility and improved experimental performance.
Error budget analysis for a single AQEC round reveals that the dominant errors are the finite coherence of the transmon qubits and residual multi-photon loss. 

Further improvements will require incorporating qubit-error-tolerant universal control of bosonic modes~\cite{Ma2020a,Reinhold2020NP,Xu2024PRX}, thereby raising gate fidelities. With such improvements, additional AQEC gates, designed via channel simulation to correct high-order photon-loss events, can be integrated into each AQEC round of our protocol.
{Moreover, our AQEC approach can be extended to higher Fock state dimensions, enabling adaption to arbitrary bosonic QEC codes~\cite{Caibosonic2021,Bosonicreview_Joshi_2021,Royer2020stab,LachanceQuirion2024} and qudit encodings~\cite{Brock2025}.}
The close agreement between the theoretical predictions and the experimental results confirms that AQEC via channel simulation provides a universal, hardware-efficient path towards fault-tolerant quantum computation. 
Together, these results mark significant progress in QEC and accelerate the development of large-scale, universal quantum computing.

\smallskip{}



%

\newpage
\bigskip
\noindent \textbf{\large{}{}Methods}{\large\par}

\noindent \textbf{Experimental devices.} The experiment is performed using the five modes highlighted in Fig.\ref{fig:concept}b (the gray region is unused). 
The storage cavity $S_1$ has the longest single-photon lifetime of $1.38\,\mathrm{ms}$ and a pure dephasing time of $6.8\,\mathrm{ms}$. 
Transmon qubit $I_1$, with an energy relaxation time of $145\,\mathrm{\mu s}$ and a pure dephasing time of $410\,\mathrm{\mu s}$, is dedicated to encode and decode the logical state in $S_1$ and to apply the photon-number-resolved a.c. Stark shift (PASS) drives to eliminate the self-Kerr effects of the storage cavity~\cite{PASS_Ma_2020}.
Transmon qubits $Y_1$ (an energy relaxation time of $135\,\mathrm{\mu s}$ and a pure dephasing time of $250\,\mathrm{\mu s}$) and $Y_2$ (an energy relaxation time of $110\,\mathrm{\mu s}$ and a pure dephasing time of $200\,\mathrm{\mu s}$) serve as ancilla qubits in Fig.~\ref{fig:concept}a. To implement the Swap gate between $Y_1$ and $Y_2$, we use cavity $S_2$ as an intermediate coupler, which has a single-photon lifetime of $1.17\,\mathrm{ms}$ and a pure dephasing time of $0.96\,\mathrm{ms}$. 

\noindent \textbf{Rate equation for AQEC.} The dynamics of our AQEC-protected quantum system is governed by a hierarchical set of coupled rate equations that describe the interplay between natural decoherence, autonomous error correction, and higher-order loss processes. For bosonic systems, the dominant error is single-photon loss with a rate $\gamma_1 = 1/T_1$, where $T_1$ is the cavity energy relaxation time. To capture the full dynamics including imperfect correction and higher-order errors, we decompose the system's Hilbert space into three subspaces with the following populations:
\begin{itemize}
\item $P_c(t)$: Population in the code subspace (quantum information intact),
\item $P_e(t)$: Population in the error subspace (single-photon loss, correctable),
\item $P_u(t)$: Population in the uncorrectable subspace (multi-photon loss, uncorrectable).
\end{itemize}
The normalization condition $P_c(t) + P_e(t) + P_u(t) = 1$ is preserved throughout the evolution. 

The free evolution of the system can be approximately described by the rate equations:
\begin{align}
\frac{dP_c}{dt} &= -\gamma_c P_c,  \\
\frac{dP_e}{dt} &= \gamma_c P_c - \gamma_e P_e, \\
\frac{dP_u}{dt} &= \gamma_e P_e.
\end{align}
Here, $\gamma_{c}$ ($\gamma_{e}$) is the single-photon loss rate in the code (error) subspace.
The evolution during time $\tau$ to second order is:
\begin{align}
P_c(t+\tau) &= P_c(t)\left(1 - \gamma_c\tau + \frac{\gamma_c^2\tau^2}{2}\right), \\
P_e(t+\tau) &= P_e(t)\left(1 - \gamma_e\tau + \frac{\gamma_e^2\tau^2}{2}\right) + \gamma_c\tau P_c(t)\left(1 - \frac{\gamma_c+\gamma_e}{2}\tau\right), \\
P_u(t+\tau) &= P_u(t) + \gamma_e\tau P_e(t) + \frac{\gamma_c\gamma_e\tau^2}{2} P_c(t).
\end{align}
The crucial term $\frac{\gamma_c\gamma_e\tau^2}{2} P_c(t)$ represents the population that undergoes two sequential decay events (Code$\rightarrow$Error$\rightarrow$Uncorrectable) within a single cycle, i.e., an irreversible loss that cannot be corrected even with perfect operations.

After the free evolution period, the correction operation of fidelity $F_{\text{success}}$ is applied, yielding:
\begin{align}
P_c^{\text{after}} &= F_{\text{success}}(P_c^{\text{before}} + P_e^{\text{before}}), \\
P_e^{\text{after}} &= 0, \\
P_u^{\text{after}} &= P_u^{\text{before}} + (1-F_{\text{success}})(P_c^{\text{before}} + P_e^{\text{before}}).
\end{align}
The key insight is that imperfect correction affects both the code subspace (even without errors) and the error subspace, injecting a fraction $(1-F_{\text{success}})$ to the uncorrectable subspace. Combining free evolution and correction, the net change in the code subspace population per cycle is:
\begin{align}
\Delta P_c &= P_c^{(n+1)} - P_c^{(n)} \\
&= P_c^{(n)}\left[F_{\text{success}}\left(1 - \gamma_c\tau + \frac{\gamma_c^2\tau^2}{2} + \gamma_c\tau - \frac{\gamma_c(\gamma_c+\gamma_e)\tau^2}{2}\right) - 1\right] \\
&= P_c^{(n)}\left[F_{\text{success}}\left(1 - \frac{\gamma_c\gamma_e\tau^2}{2}\right) - 1\right].
\end{align}
This reveals that even with perfect correction ($F_{\text{success}} = 1$), there is an irreducible loss of $\frac{\gamma_c\gamma_e\tau^2}{2}$ per cycle. For $F_{\text{success}} = 1 - \epsilon$ with $\epsilon \ll 1$, converting the discrete evolution to continuous time with $\Gamma = 1/\tau$ yields:
\begin{equation}
\frac{dP_c}{dt} = \lim_{\tau \to 0} \frac{\Delta P_c}{\tau} = -P_c\left[\frac{-\ln(F_{\text{success}})}{\tau} + \frac{\gamma_c\gamma_e\tau}{2}\right].
\end{equation}
Hence, the effective decay rate is:
\begin{equation}
{\gamma_{\text{eff}} \approx \frac{\epsilon}{\tau} + \frac{\gamma_c\gamma_e\tau}{2}},
\label{eq:EffectiveRate}
\end{equation}
where the approximation uses $-\mathrm{ln}(1-\varepsilon)\approx\varepsilon$ for $\varepsilon\ll1$.
This fundamental equation reveals the trade-off at the heart of AQEC: a higher correction rate (smaller $\tau$) reduces second-order photon losses but increases the rate of correction-induced errors.
The optimal error correction interval can then be calculated as:
\begin{equation}
{\tau_{\text{opt}} = \sqrt{\frac{2\epsilon}{\gamma_c\gamma_e}} = \sqrt{\frac{2(1-F_{\text{success}})}{\gamma_c\gamma_e}}}.
\end{equation}

Based on the fidelity after immediate encode-decode (Fig.~\ref{fig:fidelity}b), the fidelity after one round of AQEC (Fig.~\ref{fig:fidelity}d), and the intrinsic error value (Table~\ref{tab:ErrorBudget}), the average success rate of error correction can be estimated to be $F_\mathrm{success} \approx 92.4\%$. Combining this with $\gamma_c\approx\gamma_e\approx2\gamma_1$ with the natural decay rate $1/\gamma_1 \approx 1.38$ ms, we derive an optimal cycle time ${\tau_{\text{opt}}} \approx 269\,\mathrm{\mu s}$. Experimentally, we choose a free evolution time of $220 \,\mathrm{\mu s}$. The discrepancy arises from differences in the correction gate fidelity at varying free evolution durations, a factor not included in our model. 


\bigskip{}

\noindent \textbf{\large{}{}Data availability}{\large\par}

\noindent All data generated or analyzed during this study are available within the paper and its Supplementary Material. Further source data will be made available on reasonable request.

\smallskip{}

\noindent \textbf{\large{}{}Code availability}{\large\par}

\noindent The code used to solve the equations presented in the Supplementary Material will be made available on reasonable request.

\smallskip{}

\bigskip
\noindent\textbf{Acknowledgements}
This work was funded by the National Natural Science Foundation of China 
(Grants No. 12204052, 92165209, 92265210, 92365301, 92365206, 12474498, 11925404),  Innovation Program for Quantum Science and Technology (Grant No.~2021ZD0300200, 2021ZD0301800, and 2024ZD0301500). This work was also supported by the Fundamental Research Funds for the Central Universities and USTC Research Funds of the Double First-Class Initiative. This work was partially carried out at the USTC Center for Micro and Nanoscale Research and Fabrication, and  numerical calculations were performed at the Supercomputing Center of USTC.

\bigskip
\noindent\textbf{Author contributions}
L.Y.S. and C.-L.Z. supervised the project.
L.D.S. and Y.X. performed the experiment, analyzed the data, and carried out the numerical simulations. 
Y.Z. developed the FPGA technique and helped to calibrate the system.
Z.H., W.W. and J.Z. contributed to the experimental support.
W.C. fabricated the 3D cavity, conceived the protocol, and provided theoretical support. 
Z.-J.C,  L.Z.P., and Q.-X.J. contributed to the theoretical support.
X.G. and H.Y. fabricated the tantulum transmon-qubit.
Y.X., L.D.S., W.C., C.-L.Z. and L.Y.S. wrote the manuscript with feedback from all authors.

\smallskip{}

\noindent \textbf{\large{}{}Competing interests}{\large\par}

\noindent The authors declare no competing interests.

\smallskip{}

\noindent \textbf{\large{}{}Additional information}{\large\par}

\noindent \textbf{Supplementary Material} The online version contains Supplementary Material.

\noindent \textbf{Correspondence and requests for materials} should be addressed to H.Y, W.C., C.-L.Z., or L.Y.S.

\end{document}


\title{Supplementary information to: ``Extending coherence time beyond break-even point using only drives and dissipation"}

\author{Lida Sun}
\thanks{These authors contributed equally to this work.}
\author{Yifang Xu}
\thanks{These authors contributed equally to this work.}
\affiliation{Center for Quantum Information, Institute for Interdisciplinary Information Sciences, Tsinghua University, Beijing 100084, China}

\author{Yilong Zhou}
\thanks{These authors contributed equally to this work.}
\affiliation{Center for Quantum Information, Institute for Interdisciplinary Information Sciences, Tsinghua University, Beijing 100084, China}

\author{Ziyue Hua}
\affiliation{Center for Quantum Information, Institute for Interdisciplinary Information Sciences, Tsinghua University, Beijing 100084, China}

\author{Weiting Wang}
\affiliation{Center for Quantum Information, Institute for Interdisciplinary Information Sciences, Tsinghua University, Beijing 100084, China}

\author{Jie Zhou}
\affiliation{Center for Quantum Information, Institute for Interdisciplinary Information Sciences, Tsinghua University, Beijing 100084, China}

\author{Zi-Jie Chen}
\affiliation{Laboratory of Quantum Information, University of Science and Technology of China, Hefei 230026, China}

\author{Lui Zuccherelli de Paula}
\affiliation{Laboratory of Quantum Information, University of Science and Technology of China, Hefei 230026, China}

\author{Qing-Xuan Jie}
\affiliation{Laboratory of Quantum Information, University of Science and Technology of China, Hefei 230026, China}

\author{Guangming~Xue}
\affiliation{Beijing Academy of Quantum Information Sciences, Beijing 100084, China}
\affiliation{Hefei National Laboratory, Hefei 230088, China}

\author{Haifeng~Yu}
\email{hfyu@baqis.ac.cn}
\affiliation{Beijing Academy of Quantum Information Sciences, Beijing 100084, China}
\affiliation{Hefei National Laboratory, Hefei 230088, China}

\author{Weizhou Cai}
\email{caiwz@ustc.edu.cn}
\affiliation{Laboratory of Quantum Information, University of Science and Technology of China, Hefei 230026, China}

\author{Chang-Ling Zou}
\email{clzou321@ustc.edu.cn}
\affiliation{Laboratory of Quantum Information, University of Science and Technology of China, Hefei 230026, China}
\affiliation{Anhui Province Key Laboratory of Quantum Network, University of Science and Technology of China, Hefei 230026, China}
\affiliation{Hefei National Laboratory, Hefei 230088, China}

\author{Luyan Sun}
\email{luyansun@tsinghua.edu.cn}
\affiliation{Center for Quantum Information, Institute for Interdisciplinary Information Sciences, Tsinghua University, Beijing 100084, China}
\affiliation{Hefei National Laboratory, Hefei 230088, China}

\maketitle

\tableofcontents

\newpage

\renewcommand{\thefigure}{S\arabic{figure}}
\setcounter{figure}{0}
\renewcommand{\thetable}{S\arabic{table}}
\setcounter{table}{0}


\section{Experimental Device and setup}

\begin{figure}[tbp]
    \centering
    \includegraphics[width=500pt]{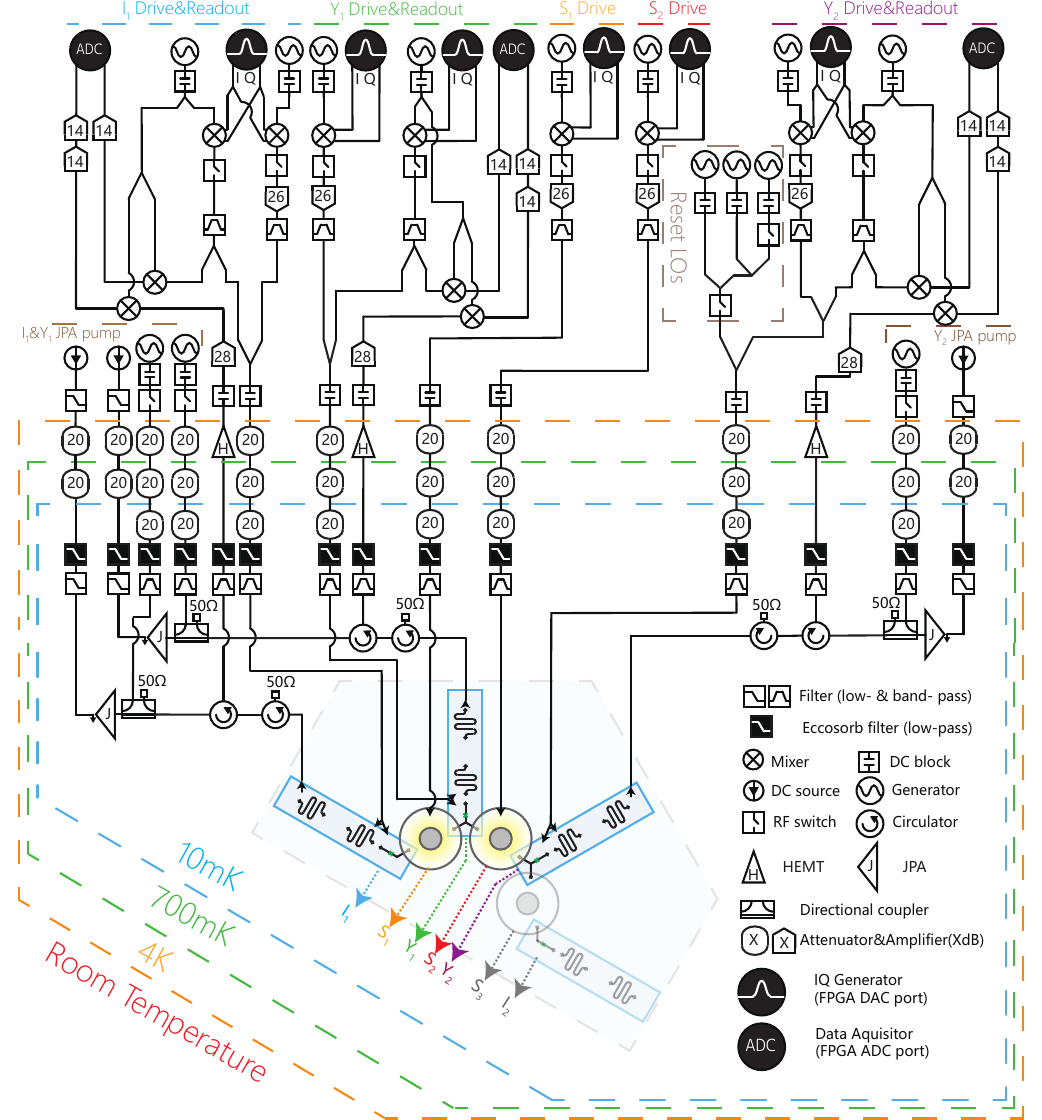}
    \caption{\textbf{Wiring diagram of the experiment.} The interpretation of the symbols in the figure are provided in the lower-right corner. The polygonal structure at the bottom represents our experimental device, which incorporates three 3D cavities (concentric circles) and four sapphire chips (blue rectangulars). The cross symbols on the chips indicate the Josephson junctions of the transmon qubits. The wiggling structures represent the readout resonators and Purcell filters. All transmon qubits, readout resonators, and coaxial cavities are controlled by IQ-modulated microwave pulses. The driving signals of each transmon qubit and the corresponding readout resonator are transmitted through the same coaxial cable, which incorporates multiple filters and attenuators to suppress noise. The transmitted readout signals are routed through two circulators to reduce back-reflections from the output port to the sample, and are amplified through Josephson parametric amplifiers (JPAs) as the first stage, followed by HEMT amplification at 4~K and room-temperature RF amplifiers. 
    }
    \label{fig:WiringDiagram}
\end{figure}

We implement the experiment in a three-dimensional (3D) circuit quantum electrodynamics (CQED) device~\cite{3Ddevice_Blais_2004, 3Ddevice_Blais_2021, 3Ddevice_Kirchmair_2013, 3Ddevice_Paik_2011, 3Ddevice_Vlastakis_2013, 3Ddevice_Wallraff_2004}. The device is similar to that used in previous works~\cite{cai2023protecting,Chen2025OpenGRAPE}, comprising three 3D coaxial stub cavities~\cite{Cav_Gao_2019, Cav_Wang_2016, 1msCav_Reagor_2016} ($S_1$, $S_2$, and $S_3$, respectively) and four superconducting transmon qubits~\cite{Transmon_Koch_2007} ($I_1$, $I_2$, $Y_1$, and $Y_2$, respectively), each equipped with a Purcell-filtered stripline readout resonator~\cite{Readout_Axline_2016, Readout_Chou_2018, ReadoutTh_Sears_2012}. The cavities are machined from a single piece of high purity (5N5) aluminum and chemically etched to enhance coherence times~\cite{Etch_Reagor_2013}. The transmon qubits and Purcell-filtered readout resonators are integrated on sapphire ($\mathrm{Al_2O_3}$) substrates, with each chip hosting one qubit and one resonator. The qubit chips are fabricated in the laboratory of the Beijing Academy of Quantum Information
Sciences (BAQIS). The $\mathrm{Al/Al_2O_3/Al}$ trilayer structures of the Josephson junctions of the transmon qubits are fabricated via electron-beam lithography and double-angle evaporation. The shunting capacitor pads and stripline readout resonators are made of tantalum in BCC alpha-phase in order to improve the qubit coherence~\cite{TaQubit_Ganjam_2024, TaQubit_Place_2021, TaQubit_Wang_2022}. The capacitor pad geometry is specifically designed for different coupling configurations: $Y_1$ and $Y_2$ couple to two cavities each, while $I_1$ and $I_2$ couple to single cavities (Fig.~\ref{fig:WiringDiagram}). The frequencies of the qubits and cavities are carefully engineered to ensure all neighboring qubit-cavity pairs operate in the dispersive coupling regime~\cite{3Ddevice_Blais_2021}.

In our experimental protocol, the fundamental mode of $S_1$ serves as the logical qubit memory. Qubit $I_1$ fulfills a dual role in quantum state encoding/decoding with $S_1$ and in implementing photon-number-resolved a.c. Stark shift (PASS) via sideband drives to mitigate Kerr-induced dephasing~\cite{PASS_Ma_2020}. Qubit $Y_1$ functions as an auxiliary element in the autonomous quantum error correction (AQEC), while the fundamental mode of $S_2$ acts as a quantum bus for excitation transfer between $Y_1$ and $Y_2$. Qubit $Y_2$ facilitates reset operations for both $S_2$ and itself. The remaining components, $S_3$ and $I_2$, are unused in this experimental configuration.

The experimental device is surrounded by a $\mu$-metal magnetic shield and installed in a dilution refrigerator with a base temperature of about 10~mK. All transmon qubits, readout resonators, and coaxial cavities are controlled by IQ-modulated microwave pulses, while the reset drives are directly output from signal generators. In this setup, we use two Sinolink SLFS0218F 5-channel signal generators, one Sinolink SLFS0218H 8-channel signal generator (three channels are used), and one Rhode-Schwartz SMB100A signal generator as the LO sources. Six synchronized BIAOZHANG PXIE-QC100 FPGA boards handle the IQ waveform generation and data acquisition, each featuring two digital-to-analog converter (DAC) ports for waveform generation, two analog-to-digital converter (ADC) ports for data acquision, and four DIO ports for RF switch control. Microwave signals are delivered through filtered coaxial cables. Readout signals are amplified through Josephson parametric amplifiers (JPAs, flux-biased by YOKOGAWA GS200 current sources) as the first stage, followed by HEMT amplification at 4~K and three room-temperature RF amplifiers. This multi-stage amplification enables high-fidelity single-shot readout of all operational qubits. Figure~\ref{fig:WiringDiagram} illustrates the detailed wiring diagram for this experiment.

\section{System Hamiltonian}
In our device, all neighboring qubits and resonators are dispersively coupled, and the system Hamiltonian can be written as:
\begin{equation}\label{eq:Ham:all}
    \hat{H}_0 = \hat{H}_\mathrm{2nd} + \hat{H}_\mathrm{4th} + \hat{H}_\mathrm{d}.
\end{equation}
Here, $\hat{H}_\mathrm{2nd}$ represents the terms that contain one pair of creation and annihilation operators, $\hat{H}_\mathrm{4th}$ represents the terms that contain two pairs of creation and annihilation operators, and $\hat{H}_\mathrm{d}$ represents the drive terms. Let $obj = \{I_1, R_{I1}, S_1, Y_1, R_{Y1}, S_2, Y_2, R_{Y2}\}$ be the name set of all experimentally used components, where $R_{I1}$, $R_{Y1}$, and $R_{Y2}$ represent the readout resonator modes of $I_1$, $Y_1$, and $Y_2$ respectively. The terms in the Hamiltonian Eq.~(\ref{eq:Ham:all}) can be explicitly written as:
\begin{equation}\label{eq:Ham:2nd}
    \hat{H}_\mathrm{2nd} =  \sum_{m\in obj}\omega_m \hat{a}_m^\dagger \hat{a}_m,
\end{equation}
\begin{equation}\label{eq:Ham:4th}
    \hat{H}_\mathrm{4th} =  \frac{1}{2}\sum_{m,n\in obj}\chi_{mn} \hat{a}_m^\dagger \hat{a}_n^\dagger \hat{a}_m \hat{a}_n,
\end{equation}
\begin{equation}\label{eq:Ham:drive}
    \hat{H}_\mathrm{d} =  \sum_{m\in obj}[\epsilon_m^\mathrm{I}*(\hat{a}_m^\dagger+\hat{a}_m)+\epsilon_m^\mathrm{Q}*i(\hat{a}_m^\dagger-\hat{a}_m)].
\end{equation}
Here, $\omega_m$ and $\chi_{mn}$ are the mode frequencies and Kerr nonlinearities, respectively, with experimentally measured values listed in Table~\ref{tab:allparameters}. $a_m$ and $a_m^\dagger$ represent the annihilation and creation operators, respectively, for mode $m$. The terms in Eq.~(\ref{eq:Ham:drive}) correspond to coherent drives. For qubits, arbitrary rotations on the Bloch sphere can be achieved by tuning the parameters $\epsilon^I$ and $\epsilon^Q$, while cavity states can undergo arbitrary displacements in the phase plane~\cite{3Ddevice_Blais_2021}. When combined with Kerr nonlinearities, the coherent drives enable more complex quantum operations, as discussed in Sec.~\ref{sec:GRAPE}.

\begin{table}
\begin{center}

\begin{tabular}{l|c|c|c|c|c|c|c|c|c} \hline \hline
 & \textbf{$I_1$} & \textbf{$R_{I1}$} & \textbf{$S_1$} & \textbf{$Y_1$} & \textbf{$R_{Y1}$} & \textbf{$S_2$} & \textbf{$Y_2$} & \textbf{$R_{Y2}$} & \textbf{$S_3$}\\ \hline

$\omega_m/2\pi \mathrm{(GHz)}$ & $4.9413$ & $7.5521$ & $6.1453$ & $5.2853$ & $7.5882$ & $5.5939$ & $4.8108$ & $7.5925$ & $6.0285$\\ 
$T_1$ ($\mu$s) & $145$ & $0.054$ & $1380$ & $135$ & $0.052$ & $1170$ & $110$ & $0.067$ & $91$\\
$T_2$ ($\mu$s) & $170$ & $-$ & $2034$ & $130$ & $-$ & $680$ & $105$ & $-$ & $-$\\
\hline \hline

 $-\chi_{ij}/2\pi \mathrm{(MHz)}$& \textbf{$I_1$} & \textbf{$R_{I1}$} & \textbf{$S_1$} & \textbf{$Y_1$} & \textbf{$R_{Y1}$} & \textbf{$S_2$} & \textbf{$Y_2$} & \textbf{$R_{Y2}$} & \textbf{$S_3$}\\ \hline

 \textbf{$I_1$} & $160$ & $2.8$ & $1.0250$ & $-$ & $-$ & $-$ & $-$ & $-$ & $-$\\
 \textbf{$R_{I1}$} & $2.8$ & $-$ & $-$ & $-$ & $-$ & $-$ & $-$ & $-$ & $-$\\
 \textbf{$S_1$} & $1.0250$ & $-$ & $0.0023$ & $0.599$ & $-$ & $0.0097$ & $\approx0$ & $-$ & $-$\\
 \textbf{$Y_1$} & $-$ & $-$ & $0.599$ & $153$ & $2.5$ & $3.091$ & $0.0158$ & $-$ & $-$\\
 \textbf{$R_{Y1}$} & $-$ & $-$ & $-$ & $2.5$ & $-$ & $-$ & $-$ & $-$ & $-$\\
 \textbf{$S_2$} & $-$ & $-$ & $0.0097$ & $3.091$ & $-$ & $0.0281$ & $0.525$ & $-$ & $-$\\
 \textbf{$Y_2$} & $-$ & $-$ & $\approx0$ & $0.0158$ & $-$ & $0.525$ & $160$ & $2.1$ & $0.25$\\
 \textbf{$R_{Y2}$} &$-$ & $-$ & $-$ & $-$ & $-$ & $-$ & $2.1$ & $-$ &$-$\\ \hline

\end{tabular}
\caption{Measured mode frequencies, coherence times (energy relaxation time $T_1$ and Ramsey dephasing time $T_2$), and Kerr nonlinearities. 
}
	\label{tab:allparameters}

\end{center}
\end{table}

\section{Coherence property and readout fidelity}
\subsection{Readout Properties}
\begin{figure}[b]
    \centering
    \includegraphics[width=400pt]{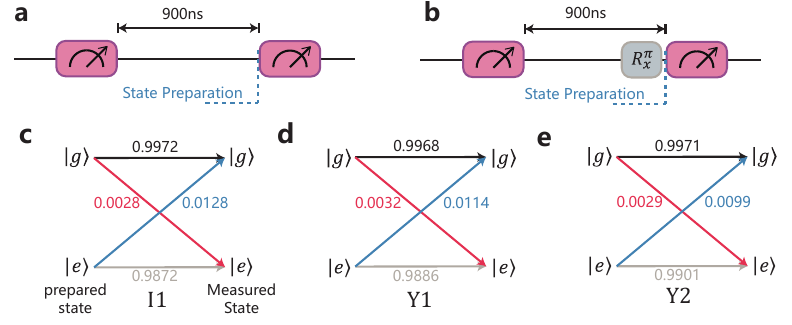}
    \caption{\textbf{Experiments to calibrate the readout properties.}
    \textbf{a, b.} Sequences for the calibration of the readout fidelity.
    \textbf{c, d, e.} Calibrated readout fidelity for qubit $I_1$, $Y_1$, and $Y_2$, respectively.
    }

    
    \label{fig:Readout properties}
\end{figure}

The readout fidelities are characterized using two sequential measurements, with the experimental sequence presented in Figs.~\ref{fig:Readout properties}a and \ref{fig:Readout properties}b. We first prepare the ground ($\ket{g}$) and excited ($\ket{e}$) states separately via post-selection and single-qubit gates, followed by readout operations. By statistically analyzing the probabilities of correct and incorrect outcomes, we extract the measurement fidelities, as shown in Figs.~\ref{fig:Readout properties}c-e. 


For the excited state $\ket{e}$, the readout infidelities ($P_\mathrm{e,g}$) are $1.28\%$, $1.14\%$, and $0.99\%$ for qubits $I_1$, $Y_1$, and $Y_2$, respectively. These errors are primarily attributed to qubit decay during the measurement process. The ground-state readout infidelities ($P_\mathrm{g,e}$) are $0.28\%$, $0.31\%$ and $0.29\%$, respectively. These errors may arise from either intrinsic error of the readout process (i.e., the readout misclassifies the input state) or the non-QND (quantum non-demolition) nature of the preceding readout, and are accounted for in later experiments such as calibration of the system's thermal population (as discussed in Sec.~\ref{sec:Thermal}). For experiments requiring high-precision state tomography, we employ Bayes' rule to mitigate the impact of readout errors.

\subsection{Coherence time, thermal populations and system purification protocol} \label{sec:Thermal}

The coherence times of each mode, including the energy relaxation time $T_1$ and the Ramsey dephasing time $T_2$, are carefully characterized, and the results are summarized in Table~\ref{tab:allparameters}. The thermal population of the qubit can be obtained by directly measuring the proportion of the excited state. Based on the analysis in the previous section, the measured excited-state population includes a contribution $P_\mathrm{g,e}$ from the ground-state readout errors rather than actual thermal excitation. Therefore, we correct the raw measurement by subtracting $P_\mathrm{g,e}$, as presented in Table~\ref{tab:Thermal}.

The thermal population of the storage cavity can be determined via Ramsey-type parity measurements~\cite{Parity_Sun_2014}. However, for a qubit simultaneously coupled to two cavities (e.g., $Y_1$ and $Y_2$), thermal excitations from both cavities interfere with the parity measurement results, making it challenging to extract the thermal population of a single cavity independently. To address this issue, we implement a chain post-selection protocol to obtain accurate measurements. Specifically, we perform simultaneous parity measurements: $I_1$ for cavity $S_1$, $Y_1$ for cavity $S_2$, and $Y_2$ for cavity $S_3$. First, the parity measurement result of $S_1$ remains unperturbed by other cavities since $I_1$ couples only to $S_1$. Second, we post-select cases where $S_1$ is in the ground state, eliminating its thermal contribution to the parity measurement of $S_2$. Finally, we further post-select cases where $S_2$ is in the ground state and evaluate the parity measurement of $S_3$. By this sequential post-selection method, we obtain reliable parity measurement results for $S_1$, $S_2$, and $S_3$. Notably, we employ this protocol to system purification in subsequent experiments: we can guarantee that all cavities and qubits remain in their ground state when all measurements yield the ground state. 

In addition to intrinsic thermal population, two key error sources affect the parity measurements: the ground-state readout error $P_\mathrm{g,e}$ and the qubit dephasing error contributing an additional population of $1/4\chi T_2$, where $\chi$ is the qubit-cavity dispersive coupling strength and $T_2$ is the dephasing time of the qubit used for the parity measurements. Table~\ref{tab:Thermal} presents the raw measurement results of the thermal populations, as well as the calculated intrinsic thermal populations after excluding ground-state readout errors and qubit dephasing effects. Note that the thermal populations of the readout resonators are estimated using the corresponding qubit $T_2$~\cite{ReadoutTh_Sears_2012}.


\begin{table}[t]
\begin{center}

\begin{tabular}{l|c|c|c|c|c|c|c|c|c} \hline \hline
 & \textbf{$I_1$} & \textbf{$R_{I1}$} & \textbf{$S_1$} & \textbf{$Y_1$} & \textbf{$R_{Y1}$} & \textbf{$S_2$} & \textbf{$Y_2$} & \textbf{$R_{Y2}$} & \textbf{$S_3$}\\ \hline

Measured population (parity measurement) & $0.66\%$ & $-$ & $0.94\%$ & $0.77\%$ & $-$ & $0.42\%$ & $0.70\%$ & $-$ & $1.29\%$\\
Contribution of $P_\mathrm{g,e}$ & $0.28\%$ & $-$ & $0.28\%$ & $0.31\%$ & $-$ & $0.31\%$ & $0.29\%$ & $-$ & $0.29\%$\\
Contribution of $T_2$ & $-$ & $-$ & $0.16\%$ & $-$ & $-$ & $0.07\%$ & $-$ & $-$ & $0.95\%$\\
Intrinsic thermal population & $0.38\%$ & $<0.1\%$ & $0.49\%$ & $0.46\%$ & $<0.1\%$ & $<0.1\%$ & $0.41\%$ & $<0.1\%$ & $<0.1\%$\\
\hline

\end{tabular}
\caption{A summary of the thermal populations in our system.}
	\label{tab:Thermal}

\end{center}
\end{table}

\section{Photon-number-resolved a.c. Stark shift}\label{sec:PASS}

Error transparency is critical for effective bosonic encoding-based error correction. Without satisfying the error transparency condition, logical states suffer dephasing when photon loss errors occur, compromising error correction. In our experiment, we employ the lowest order binomial code~\cite{BinomialTheory_Micheal_2016}, with logical states $\ket{0_\mathrm{L}} = (\ket{0}+\ket{4})/\sqrt2$ and $\ket{1_\mathrm{L} } = \ket{2}$. The error transparency condition requires $\omega_4-\omega_3 = \omega_3-\omega_1$, where $\omega_n$ denotes the frequency of the Fock $\ket{n}$ state in the storage cavity. However, the self-Kerr effect introduces nonlinearity $\omega_n = n\omega_1+K/2\cdot n(n-1)$, with $K$ being the self-Kerr coefficient. This inherently violates the error transparency condition. To address this issue, one solution is to implement the photon-number-resolved a.c. Stark shift (PASS) protocol ~\cite{PASS_Ma_2020}.

Here, we briefly outline the PASS protocol. When applying a detuned drive with strength $\Omega$ and detuning $\Delta$ to the qubit, the energy levels of the coupled cavity undergo an a.c. Stark shift. For the Fock state $\ket{n}$, the frequency shift is $-\frac{\Omega^2}{\Delta-n\chi}$. In the rotating frame, the effective frequency of the Fock $n$ energy level, under the combined effects of PASS and self-Kerr, becomes: 
\begin{equation}\begin{split}\label{eq:FreqWithPASS}
\omega_n = -\frac{\Omega^2}{\Delta-n\chi}-\frac{K}{2}n(n-1)    
\end{split}\end{equation}

In Eq.~(\ref{eq:FreqWithPASS}), we have two independently tunable parameters: $\Delta$ and $\Omega$, while the Error Transparent condition imposes one constraint, leaving one degree of freedom in the choice of parameters. An important criterion for evaluating the performance of PASS is the qubit excitation $n_\mathrm{PASS}$ induced by the detuned drive. For the Fock $n$ state, $n_\mathrm{PASS}$ can be estimated as:
\begin{equation}\begin{split}\label{eq:PASSInducedExt}
n_\mathrm{PASS} \approx \frac{\Omega^2}{(\Delta-n\chi)^2}   
\end{split}\end{equation}
Higher $n_\mathrm{PASS}$ exacerbates approximation errors and qubit decoherence, degrading PASS efficacy. 

\begin{figure}[tbp]
    \centering
    \includegraphics[width=480pt]{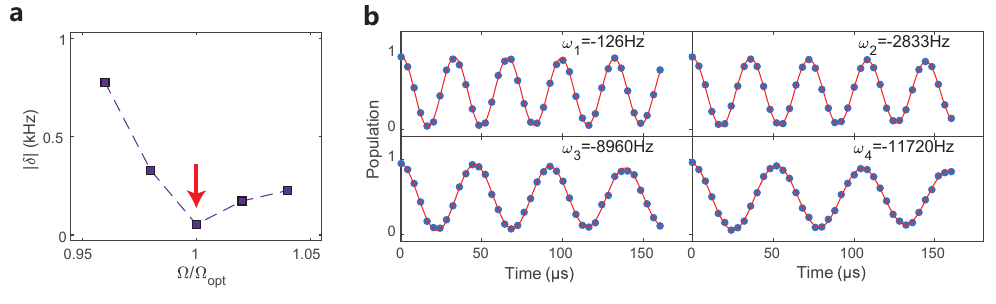}
    \caption{\textbf{Experimental calibration of PASS.} \textbf{a.} Experimentally measured absolute value of $\delta = (\omega_4-\omega_3)-(\omega_2-\omega_1)$ with respect to the drive amplitude $\Omega$ near the theoretical optimal amplitude $\Omega_\mathrm{opt}$. The detuning of the drive is fixed to $\Delta_\mathrm{opt} = -3.5\chi$. The red arrow in the figure indicates the optimal working point where the error-transparent condition is best satisfied. \textbf{b.} $\omega_1 \sim \omega_4$ at the optimal working point measured by the Ramsey experiments.
    }
    \label{fig:PASS}
\end{figure}

We optimize the optimal working point by minimize the average value of $n_{PASS}$ over Fock levels 0 to 4 while satisfying the error transparent condition. The numerically optimized result yields the best working point at $\Delta = -3.5\chi$ and $\Omega^2 = 0.4688\chi K$. Experimentally, the PASS drive is applied to $I_1$ to tune the energy levels of the logic cavity $S_1$, i.e. $\chi/2\pi = -\chi_\mathrm{I1S1}/2\pi = 1.025\,\mathrm{MHz}$ and $K/2\pi = -\chi_\mathrm{S1S1}/2\pi = 2.3\,\mathrm{kHz}$. At the predicted optimal operating point $\Delta_\mathrm{opt}/2\pi = -3.588\,\mathrm{MHz}$ and $\Omega/2\pi = 33.2\,\mathrm{kHz}$, we observe the frequency condition required for error transparency, as shown in Fig.~\ref{fig:PASS}.

\section{Quantum optimal control}\label{sec:GRAPE}
\subsection{Gradient ascent pulse engineering}

The combination of the coherent drive Hamiltonian acting on both the logical cavity ($\hat{H}_\mathrm{d}^\mathrm{s} =  \epsilon_\mathrm{s}^\mathrm{I}*(\hat{a}_\mathrm{s}^{\dagger}+\hat{a}_\mathrm{s})+\epsilon_\mathrm{ s}^\mathrm{Q}*i(\hat{a}_\mathrm{s}^\dagger-\hat{a}_\mathrm{s})$) and the qubit ($\hat{H}_\mathrm{d}^\mathrm{q} =  \epsilon_\mathrm{q}^\mathrm{I}*(\hat{a}_\mathrm{q}^{\dagger}+\hat{a}_\mathrm{q})+\epsilon_\mathrm{ q}^\mathrm{Q}*i(\hat{a}_\mathrm{q}^\dagger-\hat{a}_\mathrm{q})$) and their dispersive coupling Hamiltonian ($\hat{H}_\mathrm{disp} = -\chi \hat{a}_\mathrm{s}^{\dagger}\hat{a}_\mathrm{s}\hat{a}_\mathrm{q}^\dagger\hat{a}_\mathrm{q})$), in principle, allows universal control over the state of the logical cavity. However, for most quantum operations, finding the required drive fields ($\epsilon_\mathrm{s}^\mathrm{I}(t)$, $\epsilon_{s}^\mathrm{Q}(t)$, $\epsilon_\mathrm{q}^\mathrm{I}(t)$, $\epsilon_\mathrm{q}^\mathrm{Q}(t)$ is generally non-trivial. 

GRadient Ascent Pulse Engineering (GRAPE) ~\cite{GRAPE_Khaneja_2005, GRAPE_Chen_2025} provides a powerful numerical method to determine the control fields. Given the initial and target states of the quantum operation and the system Hamiltonian, GRAPE iteratively optimizes the control pulses ($\epsilon_\mathrm{s}^\mathrm{I}(t)$, $\epsilon_\mathrm{s}^\mathrm{Q}(t)$, $\epsilon_\mathrm{q}^\mathrm{I}(t)$, $\epsilon_\mathrm{q}^\mathrm{Q}$(t)) to minimize the infidelity between the achieved final state and the target state, thereby realizing the desired quantum operation. To enable numerical computation, we discretize the time interval into small steps $t_1 ,t_2,..., t_N$, where $t_N = T$ represents the total duration of the quantum operation. The GRAPE objective function can then be formulated as:
\begin{equation}\label{eq:GRAPEObjFunc}
    \Phi_0 =  1-\frac{1}{m}\sum_{i=1}^{m}{\left|\bra{\psi_i^\mathrm{f}}\hat{U}_N \hat{U}_{N-1}... \hat{U}_1 \ket{\psi_i^0}\right|^2}
\end{equation}
where $\ket{\psi_i^0}$ and $\ket{\psi_i^\mathrm{f}}$ are the $i$-th initial state and the target state for the desired quantum operation, respectively. $\hat{U}_{i}$ are the time evolution operators defined as:
\begin{equation}\label{eq:TimeEvoOper}
    \hat{U}_{i} =  \exp{[-i\hat{H}(t_i)(t_i-t_{i-1})]}.
\end{equation}
Here, $\hat{H}(t_i)$ represents the total system Hamiltonian at time $t_i$. The gradient of the objective function can also be efficiently calculated during each iteration ~\cite{GRAPE_Khaneja_2005, GRAPE_Chen_2025}. To address the limitations of real-world microwave sources in terms of amplitude and bandwidth, we aim to avoid excessively large amplitudes or sharp variations in the final control waveform. To enforce this constraint, we introduce an additional penalty term into the objective function as follows:
\begin{equation}\label{eq:GRAPEShapePunish}
    \Phi_\mathrm{shape} =  \frac{\alpha}{N}\sum_x{(\sum_{i=1}^{N}{(\exp{(\frac{\epsilon_x(t_i)}{h})^2}-1)}+\sum_{i=1}^{N}{(\exp{(\frac{\epsilon_x(t_i)-\epsilon_x(t_{i-1})}{h_\mathrm{d}})^2}-1)})}.
\end{equation}
Note that $\Phi_\mathrm{shape}$ is summed over all $\epsilon_x\in\{\epsilon_\mathrm{s}^\mathrm{I}, \epsilon_\mathrm{s}^\mathrm{Q}, \epsilon_\mathrm{q}^\mathrm{I}, \epsilon_\mathrm{q}^\mathrm{Q}\}$. Here, $\alpha$ represents the tradeoff between state fidelity and shape feasibility, which is chosen to be {$\alpha = 0.1/(number~of~\epsilon$)} in our experiment; $h$ and $h_\mathrm{d}$ represent the penalty coefficients for the amplitude and the variation of the control waveform, respectively. We choose $h=480\,\mathrm{MHz}$ and $h_\mathrm{d}=15\,\mathrm{MHz}$ in the experiment considering the ability of our microwave source. The value and gradient of $\Phi_\mathrm{shape}$ can also be easily calculated from the shape of the current pulse. In conclusion, our goal is to minimize $\Phi_\mathrm{tot} = \Phi_0 + \Phi_\mathrm{shape}$.

In our experiments, the GRAPE algorithm is restricted to operate solely on the lowest two energy levels (i.e. $\ket{g}$ and $\ket{e }$) of the qubit. Accordingly, the qubit-related operators are truncated to two-level Pauli operators, that is:
\begin{equation}
    \hat{a}_\mathrm{q}\rightarrow\hat{\sigma}_-, \hat{a}_\mathrm{q}^{\dagger}\rightarrow\hat{\sigma}_+
\end{equation}

\subsection{Encode and decode}
The encode and decode operations refer to the mapping between the transmon qubit state and the logical state in the cavity. These operations enable initialization and readout of the logical states. In our experiment, we implement encode and decode operations using $I_1$ for cavity $S_1$. We denote $\ket{0_\mathrm{L}} = (\ket{0}+\ket{4})/\sqrt{2}$ and $\ket{1_\mathrm{L} } = \ket{2}$ as the unperturbed logical states of the lowest-order binomial code, and $\ket{L_\mathrm{dual}} = (\ket{0}-\ket{4})/\sqrt{2}$ which provides partial compensation for the ancilla-induced dephasing errors (see Sec.~\ref{sec:AQECPulse}). In the joint Hilbert space $I_1 \otimes S_1$, the initial and target states of the encode and decode processes can be expressed as:
\begin{equation}\label{eq:Encode}
\mathrm{Encode}: 
\begin{cases} 
\ket{g,0}\rightarrow\ket{g,0_\mathrm{L}} \\
\ket{e ,0}\rightarrow\ket{g,1_\mathrm{L} } \\
(\ket{g,0}+\ket{e ,0})/\sqrt2\rightarrow(\ket{g,0_\mathrm{L}}+\ket{g,1_\mathrm{L} })/\sqrt2 \\
(\ket{g,0}-\ket{e ,0})/\sqrt2\rightarrow(\ket{g,0_\mathrm{L}}-\ket{g,1_\mathrm{L} })/\sqrt2 \\
(\ket{g,0}+\mathrm{i}\cdot\ket{e ,0})/\sqrt2\rightarrow(\ket{g,0_\mathrm{L}}+\mathrm{i}\cdot\ket{g,1_\mathrm{L} })/\sqrt2 \\
(\ket{g,0}-\mathrm{i}\cdot\ket{e ,0})/\sqrt2\rightarrow(\ket{g,0_\mathrm{L}}-\mathrm{i}\cdot\ket{g,1_\mathrm{L} })/\sqrt2, 
\end{cases}
\end{equation}

\begin{equation}\label{eq:DecodeWithDual}
\mathrm{Decode}: 
\begin{cases} 
\ket{g,0_\mathrm{L}}\rightarrow\ket{g,0} \\
\ket{g,1_\mathrm{L} }\rightarrow\ket{e ,0} \\
(\ket{g,0_\mathrm{L}}+\ket{g,1_\mathrm{L} })/\sqrt2\rightarrow(\ket{g,0}+\ket{e ,0})/\sqrt2 \\
(\ket{g,0_\mathrm{L}}-\ket{g,1_\mathrm{L} })/\sqrt2\rightarrow(\ket{g,0}-\ket{e ,0})/\sqrt2 \\
(\ket{g,0_\mathrm{L}}+i\cdot\ket{g,1_\mathrm{L} })/\sqrt2\rightarrow(\ket{g,0}+i\cdot\ket{e ,0})/\sqrt2 \\
(\ket{g,0_\mathrm{L}}-i\cdot\ket{g,1_\mathrm{L} })/\sqrt2\rightarrow(\ket{g,0}-i\cdot\ket{e ,0})/\sqrt2 \\
\ket{g,L_\mathrm{dual}}\rightarrow\ket{g,1}. 
\end{cases}
\end{equation}

While Eq.~\ref{eq:Encode} shows that the first and third conditions are sufficient to derive the remaining four conditions mathematically, we include all six Bloch sphere poles explicitly in the GRAPE optimization for two important reasons: First, numerical optimization methods may produce control pulses that perform well only on the specified conditions but relatively poorly on the implicitly derived ones. Second, explicitly incorporating all six conditions ensures the resulting pulses maintain robust performance across the entire Hilbert space.
For the Encode and Decode operations, we work within the truncated subspace of $I_1 \otimes S_1$, described by the Hamiltonian:
\begin{equation}\label{eq:Ham:Encode&Decode}
\begin{gathered}
    \hat{H}_\mathrm{En/De} =  \epsilon_\mathrm{S1}^\mathrm{I}*(\hat{a}_\mathrm{S1}^{\dagger}+\hat{a}_\mathrm{S1})+\epsilon_\mathrm{S1}^\mathrm{Q}*i(\hat{a}_\mathrm{S1}^\dagger-\hat{a}_\mathrm{S1})
    +\epsilon_\mathrm{I1}^\mathrm{I}*\hat{\sigma}_{x\mathrm{I1}}+\epsilon_\mathrm{I1}^\mathrm{Q}*\hat{\sigma}_{y\mathrm{I1}}\\
    +\chi_\mathrm{I1S1}\hat{a}_\mathrm{S1}^{\dagger}\hat{a}_\mathrm{S1}\ket{e _\mathrm{I1}}\bra{e _\mathrm{I1}}
    +\frac{\chi_\mathrm{S1S1}}{2}\hat{a}_\mathrm{S1}^{\dagger}\hat{a}_\mathrm{S1}^{\dagger}\hat{a}_\mathrm{S1}\hat{a}_\mathrm{S1}.
\end{gathered}
\end{equation}

\begin{figure}[t]
    \centering
    \includegraphics[width=480pt]{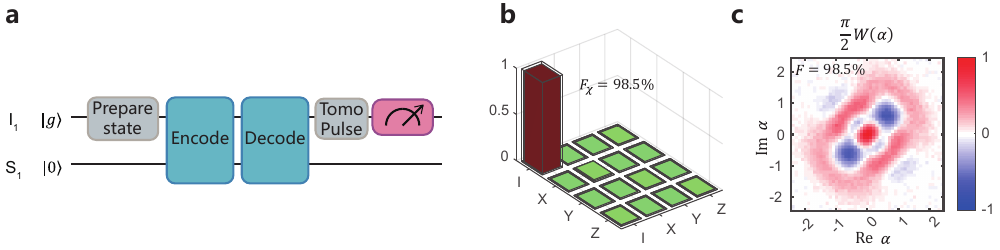}
    \caption{\textbf{Experimental sequence and results of calibration of the Encode-Decode process.}
    \textbf{a.} Sequence for measuring the joint process fidelity of the Encode and Decode process. 
    \textbf{b.} Measured process matrix. \textbf{c.} Measured Wigner function of the logical state $\ket{0_\mathrm{L}}-i\cdot\ket{1_\mathrm{L}}$ prepared by the Encode pulse. 
    }
    \label{fig:En_De_SI}
\end{figure}

The Encode and Decode operations are implemented with durations of $1.2\,\mathrm{\mu s}$ and $1.8\,\mathrm{\mu s}$, respectively. We use the GRAPE method to generate the control pulses and experimentally measure the process fidelity of the joint process containing one Encode and one Decode in succession, as shown in Fig.~\ref{fig:En_De_SI}a. The combined process fidelity is $98.5\%$, with the corresponding process matrix shown in Fig.~\ref{fig:En_De_SI}b. Additionally, we perform Wigner tomography on the logical state $(\ket{0_\mathrm{L}}-\mathrm{i}\cdot\ket{1_\mathrm{L} })/\sqrt{2}$ prepared by the Encode pulse, obtaining a state fidelity of $98.5\%$ (Fig.~\ref{fig:En_De_SI}c).

\subsection{Swap the excitation}
Transferring the excitation from qubit $Y_1$ to qubit $Y_2$ significantly reduces the dephasing of the logical state in cavity $S_1$ caused by the reset operation. This constitutes a key characteristic of our work. We implement this transfer operation using cavity $S_2$ as an intermediary bus. The desired transformation in the joint Hilbert space $Y_1\otimes S_2\otimes Y_2$ can be expressed as:
\begin{equation}\label{IntuitiveSwapCondition}
\mathrm{Ideal\ Excitation\ Transfer:} 
\begin{cases} 
\ket{g,0,g}\rightarrow\ket{g,0,g} \\
\ket{e ,0,g}\rightarrow\ket{g,0,e }. 
\end{cases}
\end{equation}

However, the logical state remains present in cavity $S_1$ during the transfer. Owing to the dispersive coupling between $Y_1$ and $S_1$, the presence of a finite population in $S_1$ leads to a frequency shift in $Y_1$'s resonance relative to its isolated value. Taking this effect into account, we adopt a generalized set of constraints (in the joint Hilbert space $S_1 \otimes Y_1 \otimes S_2 \otimes Y_2$) to ensure robustness under realistic experimental conditions:
\begin{equation}\label{RealSwapCondition}
\mathrm{Real\ Excitation\ Transfer}: 
\begin{cases} 
\ket{0,g,0,g}\rightarrow\ket{0,g,0,g} \\
\ket{0,e ,0,g}\rightarrow\ket{0,g,0,e } \\
\ket{1,g,0,g}\rightarrow\ket{1,g,0,g} \\
\ket{1,e ,0,g}\rightarrow\ket{1,g,0,e } \\
\ket{2,g,0,g}\rightarrow\ket{2,g,0,g} \\
\ket{2,e ,0,g}\rightarrow\ket{2,g,0,e } \\
\ket{3,g,0,g}\rightarrow\ket{3,g,0,g} \\
\ket{3,e ,0,g}\rightarrow\ket{3,g,0,e } \\
\ket{4,g,0,g}\rightarrow\ket{4,g,0,g} \\
\ket{4,e ,0,g}\rightarrow\ket{4,g,0,e }. 
\end{cases}
\end{equation}

These state transformations can be achieved through coherent control pulses applied to $Y_1$, $S_2$, and $Y_2$, combined with the dispersive coupling of the system. Note that the drive on $S_1$ is not necessary, since the conditions do not involve direct transitions between $S_1$'s energy levels. This feature allows us to truncate cavity $S_1$ to its first five levels during GRAPE optimization, significantly reducing the computational resource overhead. Therefore, we simplify the system Hamiltonian by including only the relevant drive terms on $Y_1$, $S_2$, and $Y_2$ when projecting onto the target subspace. The resulting effective Hamiltonian is given by: 
\begin{equation}\label{eq:Ham:Swap}
\begin{gathered}
    \hat{H}_\mathrm{swap} =  \epsilon_\mathrm{Y1}^\mathrm{I}*\hat{\sigma}_{x\mathrm{Y1}}+\epsilon_\mathrm{Y1}^\mathrm{Q}*\hat{\sigma}_{y\mathrm{Y1}}+\epsilon_{S2}^\mathrm{I}*(\hat{a}_\mathrm{S2}^{\dagger}+\hat{a}_\mathrm{S2})+\epsilon_\mathrm{S2}^\mathrm{Q}*i(\hat{a}_\mathrm{S2}^\dagger-\hat{a}_\mathrm{S2})
    +\epsilon_\mathrm{Y2}^\mathrm{I}*\hat{\sigma}_{x\mathrm{Y2}}+\epsilon_\mathrm{Y2}^\mathrm{Q}*\hat{\sigma}_{y\mathrm{Y2}}\\
    +\chi_\mathrm{Y1S1}\hat{a}_\mathrm{S1}^{\dagger}\hat{a}_\mathrm{S1}\ket{e _\mathrm{Y1}}\bra{e _\mathrm{Y1}}
    +\chi_\mathrm{Y1S2}\hat{a}_\mathrm{S2}^{\dagger}\hat{a}_\mathrm{S2}\ket{e _\mathrm{Y1}}\bra{e _\mathrm{Y1}}
    +\chi_\mathrm{Y2S2}\hat{a}_\mathrm{S1}^{\dagger}\hat{a}_\mathrm{S1}\ket{e _\mathrm{Y2}}\bra{e _\mathrm{Y2}}\\
    +\frac{\chi_\mathrm{S1S1}}{2}\hat{a}_\mathrm{S1}^{\dagger}\hat{a}_\mathrm{S1}^{\dagger}\hat{a}_\mathrm{S1}\hat{a}_\mathrm{S1}
    +\frac{\chi_\mathrm{S2S2}}{2}\hat{a}_\mathrm{S2}^{\dagger}\hat{a}_\mathrm{S2}^{\dagger}\hat{a}_\mathrm{S2}\hat{a}_\mathrm{S2}
    +\chi_\mathrm{S1S2}\hat{a}_\mathrm{S1}^{\dagger}\hat{a}_\mathrm{S1}\hat{a}_\mathrm{S2}^{\dagger}\hat{a}_\mathrm{S2}
    +\chi_\mathrm{Y1Y2}\ket{e _\mathrm{Y1}}\bra{e _\mathrm{Y1}}\ket{e _\mathrm{Y2}}\bra{e _\mathrm{Y2}}.
\end{gathered}
\end{equation}

\begin{figure}[tbp]
    \centering
    \includegraphics{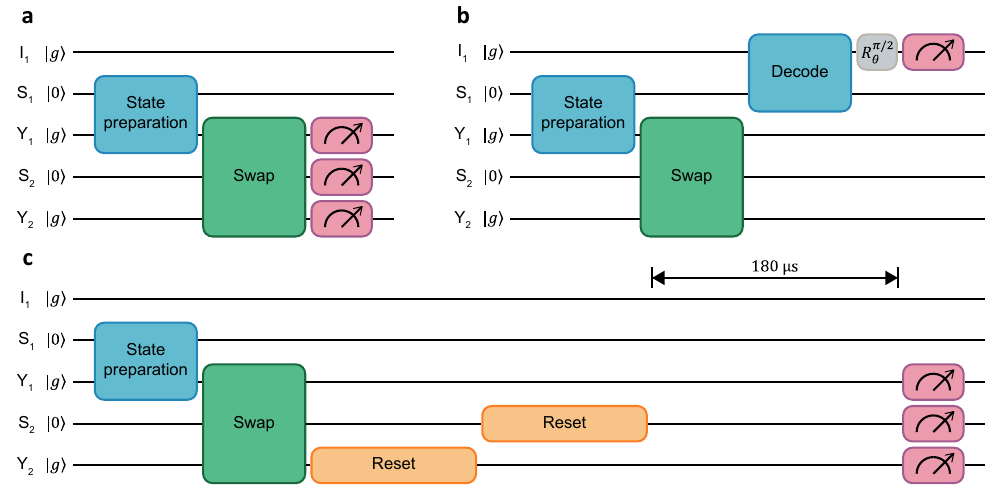}
    \caption{\textbf{Sequences for different calibration experiments.}
    \textbf{a.} Characterize the performance of the Swap pulse via residual population in $Y_1$, $S_2$, and $Y_2$. Populations on $S_2$ are measured from parity measurement using $Y_2$.
    \textbf{b.} Calibrate the induced phase on Fock states in $S_1$ by the Swap pulse.
    \textbf{c.} Calibrate the residual population in $Y_1$, $S_2$, and $Y_2$ after reset and a long free-evolution. Populations on $S_2$ are measured from parity measurement using $Y_2$.
    }
    \label{fig:CalSwapCircuit}
\end{figure}

The Swap gate operation time is implemented with a duration of $1.6\,\mathrm{\mu s}$ using GRAPE-optimized control pulses. To characterize the gate performance, we prepare $S_1$ in Fock states $\ket{0}$ through $\ket{4}$, while initializing $Y_1$ in either the ground state $\ket{g}$ or excited state $\ket{e }$. After applying a single Swap operation, we measure the residual population on $Y_1$, $S_2$, and $Y_2$, where the residual population on $S_2$ is obtained by parity measurement assisted by $Y_2$. The sequence for this experiment is shown in Fig.~\ref{fig:CalSwapCircuit}a, and the results are summarized in Table~\ref{tab:ResidualPop}. When $Y_1$ is prepared in the excited state, the resulting residual population of $Y_1$ is close to 0, while that of $Y_2$ is close to 1, indicating the success of the excitation transfer operation.

\begin{table}
\begin{center}
\begin{tabular}{c|c|c|c|c|c} \hline \hline
 \textbf{Initial state on $Y_1$ and $S_1$} & $S_1$:$\ket{0}$ & $S_1$:$\ket{1}$ & $S_1$:$\ket{2}$ & $S_1$:$\ket{3}$ & $S_1$:$\ket{4}$ \\ \hline
 \multirow{3}{*}{$Y_1$:$\ket{g}$} 
& $Y_1$: $0.55\%$ & $Y_1$: $0.72\%$ & $Y_1$: $1.09\%$ & $Y_1$: $0.92\%$ & $Y_1$: $1.43\%$ \\
& $S_2$: $0.73\%$ & $S_2$: $0.77\%$ & $S_2$: $0.95\%$ & $S_2$: $0.82\%$ & $S_2$: $0.92\%$ \\
& $Y_2$: $2.15\%$ & $Y_2$: $1.98\%$ & $Y_2$: $3.30\%$ & $Y_2$: $3.67\%$ & $Y_2$: $4.03\%$ \\ \hline
 \multirow{3}{*}{$Y_1$:$\ket{e}$} 
& $Y_1$: $0.67\%$ & $Y_1$: $0.63\%$ & $Y_1$: $1.13\%$ & $Y_1$: $1.24\%$ & $Y_1$: $1.30\%$ \\
& $S_2$: $2.78\%$ & $S_2$: $2.86\%$ & $S_2$: $2.68\%$ & $S_2$: $2.76\%$ & $S_2$: $2.97\%$ \\
& $Y_2$: $96.82\%$ & $Y_2$: $95.83\%$ & $Y_2$: $95.36\%$ & $Y_2$: $95.56\%$ & $Y_2$: $93.19\%$ \\ \hline
\hline
\end{tabular}
\caption{\textbf{Measured residual populations after the Swap operation.} Measured population on $Y_1$, $S_2$, and $Y_2$ at the end of the sequence illustrated in Fig.~\ref{fig:CalSwapCircuit}a. The population on $S_2$ is obtained by parity measurement assisted by $Y_2$.
	\label{tab:ResidualPop}
}
\end{center}
\end{table}

\subsection{Swap-induced phase}
During the excitation transfer process, $Y_1$ retains a finite population. Due to the dispersive coupling between $Y_1$ and $S_1$, this population induces Fock-state-dependent phase accumulations in $S_1$. We denote the phase introduced by the Swap pulse on Fock state $\ket{n}$ of $S_1$ when $Y_1$ is initially in the ground state as $\phi_{gn}$, and the phase introduced when $Y_1$ is initially in the excited state as $\phi_{e n}$. According to our AQEC protocol (see Fig.~\ref{fig:Evolution_Map_SI}), we aim to preserve unperturbed logical states in $S_1$ after the excitation transfer. Consequently, we need to calibrate these phase shifts and compensate them in the AQEC procedure (see Sec.~\ref{sec:AQECPulse}). 

The experimental sequence for this phase calibration is illustrated in Fig.~\ref{fig:CalSwapCircuit}b. We first prepare the $\ket{g}$ or $\ket{e}$ states in $Y_1$, along with the superposition states $\ket{0}+\ket{n} (n=0\sim4)$ in $S_1$. Then we apply the Swap pulse, which transforms $\ket{0}+\ket{n}$ in $S_1$ to $\ket{0}+\exp(i\phi_\mathrm{swap})\ket{n}$, where $\phi_\mathrm{swap} = \phi_{gn}$ or $\phi_{en}$ depends on the initial state. Subsequently, we decode the resulting $\ket{0}+\exp(i\phi_\mathrm{swap})\ket{n}$ state to $\ket{g}+\exp(i\phi_\mathrm{swap})\ket{e }$ in $I_1$ via a decoding pulse governed by the transformation conditions specified as (on the joint Hilbert space $I_1\otimes S_1$):
\begin{equation}\label{DecodeThePhaseToQubit}
\mathrm{Decode\ The\ Phase\ To\ Qubit:} 
\begin{cases} 
\ket{g,0}\rightarrow\ket{g,0} \\
\ket{g,n}\rightarrow\ket{e ,0} \\
(\ket{g,0}+\ket{g,n})/\sqrt2\rightarrow(\ket{g,0}+\ket{e ,0})/\sqrt2 \\
(\ket{g,0}-\ket{g,n})/\sqrt2\rightarrow(\ket{g,0}-\ket{e ,0})/\sqrt2 \\
(\ket{g,0}+i\cdot\ket{g,n})/\sqrt2\rightarrow(\ket{g,0}+i\cdot\ket{e ,0})/\sqrt2 \\
(\ket{g,0}-i\cdot\ket{g,n})/\sqrt2\rightarrow(\ket{g,0}-i\cdot\ket{e ,0})/\sqrt2.
\end{cases}
\end{equation}

\begin{figure}[tbp]
    \centering
    \includegraphics[width=480pt]{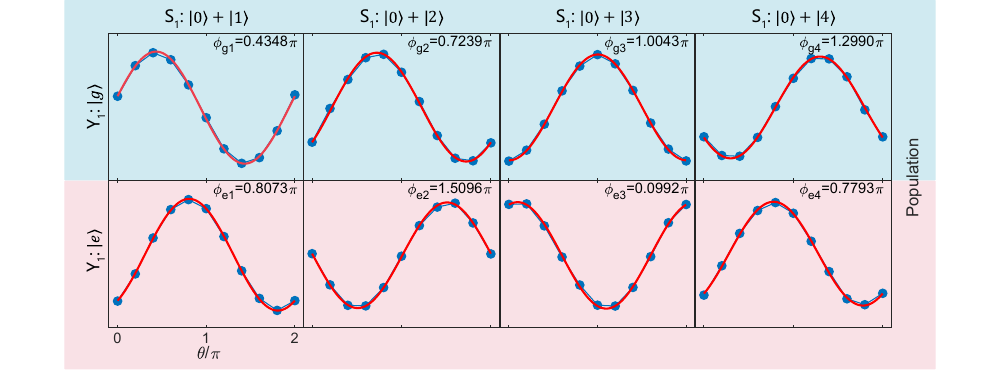}
    \caption{\textbf{Experimental calibration result of the Swap-induced phases $\phi_{gn}$ and $\phi_{e n}$.}
    Measured population on $I_1$ with respect to $\theta$ at the end of the sequence illustrated in Fig.~\ref{fig:CalSwapCircuit}b, when $Y_1$ and $S_1$ are initialized in different initial states, respectively. The data exhibit a cosine dependence, from which $\phi_{gn}$ and $\phi_{e n}$ are extracted through curve fitting.
    }
    \label{fig:SwapInducedPhase}
\end{figure}

Finally, we apply a $\pi/2$ rotation $\hat{R}_\theta (\pi/2)$ to qubit $I_1$ with $\theta \in [0,2\pi]$. The resulting population of $I_1$ follows $P_\mathrm{I1}=(1+\cos(\theta-\phi_\mathrm{swap}))/2$. The value of $\phi_\mathrm{swap}$ can be determined by cosine fitting of the data. The experimentally obtained results are presented in Fig.~\ref{fig:SwapInducedPhase}.

\subsection{Autonomous quantum error correction}\label{sec:AQECPulse}
The AQEC pulse must satisfy two key requirements: 1) map states in the error subspace back to their corresponding positions in the logical subspace; 2) leave states initially in the logical subspace unaffected. To preserve the unitarity of the quantum operation, the AQEC pulse excites the ancilla when acting on error states, enabling logical state restoration through subsequent excitation transfer and reset operations. 

Accounting for both phase accumulation at each energy level of the logical cavity during the free evolution process and the influence of no-jump errors~\cite{BinomialTheory_Micheal_2016}, the states in the logical and error subspaces immediately prior to the AQEC pulse differ from their ideal forms. The actual states are given by:
\begin{equation}\label{LogicalStatesBeforeAQEC}
\mathrm{Logical\ States\ Before\ AQEC:} 
\begin{cases} 
\ket{0_\mathrm{L}^1} = \frac{1}{\sqrt{1+\exp(-4\kappa t_\mathrm{FE})}}(\ket{0}+\exp(-2\kappa t_\mathrm{FE}-i\omega_4 t_\mathrm{FE})\ket{4})\\
\ket{1_\mathrm{L} ^1} = \exp(-i\omega_2 t_\mathrm{FE})\ket{2},
\end{cases}
\end{equation}

\begin{equation}\label{ErrorStatesBeforeAQEC}
\mathrm{Error\ States\ Before\ AQEC:} 
\begin{cases} 
\ket{0_\mathrm{E}^1} = \exp(-i\omega_3 t_\mathrm{FE})\ket{3}\\
\ket{1_\mathrm{E} ^1} = \exp(-i\omega_1 t_\mathrm{FE})\ket{1}.
\end{cases}
\end{equation}
where $\kappa = 1/(1380\,\mathrm{\mu s})$ is the logical cavity decay rate and $t_\mathrm{FE} = 220\,\mu s$ is the free evolution time in one AQEC cycle. Note that the phase coefficients preceding each state cannot be neglected, as they play a crucial role in the superposition states. To restore the logical subspace to its ideal form after the excitation transfer, we implement active phase compensation that reverses the Swap-induced phase accumulation. The resulting logical state after AQEC is defined by:
\begin{equation}\label{FromLogicalSpace}
\mathrm{From\ Logical\ Subspace:} 
\begin{cases} 
\ket{0_\mathrm{L}^2} = \frac{1}{\sqrt{2}}\cdot(\ket{0}+\exp(-i\phi_{4g})\ket{4})\\
\ket{1_\mathrm{L}^2} = \exp(-i\phi_{2g})\ket{2},
\end{cases}
\end{equation}
\begin{equation}\label{FromErrorSpace}
\mathrm{From\ Error\ Subspace:} 
\begin{cases} 
\ket{0_\mathrm{L}^{2*}} = \frac{1}{\sqrt{2}}\cdot(\ket{0}+\exp(-i\phi_{4e})\ket{4})\\
\ket{1_\mathrm{L}^{2*}} = \exp(-i\phi_{2e})\ket{2}.
\end{cases}
\end{equation}

When the logical state is initially in either the logical space or the error space, the final states are mapped to the spaces spanned by $\{\ket{0_\mathrm{L}^2},\ket{1_\mathrm{L}^2}\}$ and $\{\ket{0_\mathrm{L}^{2*}}, \ket{1_\mathrm{L}^{2*}}\}$, respectively. This distinction arises because the Swap pulse introduces different phases ($\phi_{gn}\neq\phi_{en}$) on the logical state depending on whether the ancilla state is initially in the ground state or the excited state. Both logical spaces must return to the unperturbed logical space after the excitation transfer process. Experimentally, we use $Y_1$ as the ancilla. We therefore define the transformation conditions (on the joint Hilbert space $Y_1\otimes S_1$) for the AQEC pulse as follows:
\begin{equation}\label{eq:AQECWithDual}
\mathrm{AQEC:} 
\begin{cases} 
\ket{g,0_\mathrm{L}^1}\rightarrow\ket{g,0_\mathrm{L}^2} \\
\ket{g,1_\mathrm{L}^1}\rightarrow\ket{g,1_\mathrm{L}^2} \\
(\ket{g,0_\mathrm{L}^1}+\ket{g,1_\mathrm{L}^1})/\sqrt2\rightarrow(\ket{g,0_\mathrm{L}^2}+\ket{g,1_\mathrm{L}^2})/\sqrt2 \\
(\ket{g,0_\mathrm{L}^1}-\ket{g,1_\mathrm{L}^1})/\sqrt2\rightarrow(\ket{g,0_\mathrm{L}^2}-\ket{g,1_\mathrm{L}^2})/\sqrt2 \\
(\ket{g,0_\mathrm{L}^1}+i\cdot\ket{g,1_\mathrm{L}^1})/\sqrt2\rightarrow(\ket{g,0_\mathrm{L}^2}+i\cdot\ket{g,1_\mathrm{L}^2})/\sqrt2 \\
(\ket{g,0_\mathrm{L}^1}-i\cdot\ket{g,1_\mathrm{L}^1})/\sqrt2\rightarrow(\ket{g,0_\mathrm{L}^2}-i\cdot\ket{g,1_\mathrm{L}^2})/\sqrt2 \\
\ket{g,0_\mathrm{E}^1}\rightarrow\ket{e,0_\mathrm{L}^{2*}} \\
\ket{g,1_\mathrm{E}^1}\rightarrow\ket{e,1_\mathrm{L}^{2*}} \\
(\ket{g,0_\mathrm{E}^1}+\ket{g,1_\mathrm{E}^1})/\sqrt2\rightarrow(\ket{e,0_\mathrm{L}^{2*}}+\ket{e,1_\mathrm{L}^{2*}})/\sqrt2 \\
(\ket{g,0_\mathrm{E}^1}-\ket{g,1_\mathrm{E}^1})/\sqrt2\rightarrow(\ket{e,0_\mathrm{L}^{2*}}-\ket{e,1_\mathrm{L}^{2*}})/\sqrt2 \\
(\ket{g,0_\mathrm{E}^1}+i\cdot\ket{g,1_\mathrm{E}^1})/\sqrt2\rightarrow(\ket{e,0_\mathrm{L}^{2*}}+i\cdot\ket{e,1_\mathrm{L}^{2*}})/\sqrt2 \\
(\ket{g,0_\mathrm{E}^1}-i\cdot\ket{g,1_\mathrm{E}^1})/\sqrt2\rightarrow(\ket{e,0_\mathrm{L}^{2*}}-i\cdot\ket{e,1_\mathrm{L}^{2*}})/\sqrt2 \\
\ket{g,L_\mathrm{dual}^1}\rightarrow\ket{g,L_\mathrm{dual}^2},
\end{cases}
\end{equation}
where $\ket{L_\mathrm{dual}^1}$ and $\ket{L_\mathrm{dual}^2}$ are defined as:
\begin{equation}\label{InitDual} 
\ket{L_\mathrm{dual}^1} =  \frac{1}{\sqrt{1+\exp(-4\kappa t_\mathrm{FE})}}(\exp(-2\kappa t_\mathrm{FE})\ket{0}-\exp(-i\omega_4 t_\mathrm{FE})\ket{4}),
\end{equation}
\begin{equation}\label{FinDual} 
\ket{L_\mathrm{dual}^2} =\frac{1}{\sqrt{2}}(\ket{0}-\exp(-i\phi_{4g})\ket{4}).
\end{equation}

These two states are introduced to enhance the code's robustness against cavity dephasing. During the free evolution process, thermal excitations from $I_1$ and $S_1$ can completely randomize the relative phases among the Fock states in the $S_1$ cavity. By combining the first and last conditions in Eq.~\ref{eq:AQECWithDual}, it follows that if the state of cavity $S_1$ initially resides in the subspace $\{\ket{0},\ket{4}\}$ prior to the AQEC pulse, the AQEC procedure will preserve the state within this subspace, regardless of the relative phase between $\ket{0}$ and $\ket{4}$. Furthermore, by incorporating the first and last conditions in Eq.~\ref{eq:DecodeWithDual}, if the state in $S_1$ resides in the subspace $\{\ket{0},\ket{4}\}$ before decoding, the final decoded state of $I_1$ must be $\ket{g}$. With these additional constraints, the influence of logical cavity dephasing on $\ket{0_\mathrm{L}}$ can be effectively eliminated. In our experiment, the AQEC process duration is set to $4.4\,\mu$s. The system Hamiltonian is projected onto the $Y_1\otimes S_1$ subspace, as defined by:
\begin{equation}\label{eq:Ham:Encode&Decode}
\begin{gathered}
    \hat{H}_\mathrm{EnDe} =  \epsilon_\mathrm{S1}^\mathrm{I}*(\hat{a}_\mathrm{S1}^{\dagger}+\hat{a}_\mathrm{S1})+\epsilon_\mathrm{S1}^\mathrm{Q}*i(\hat{a}_\mathrm{S1}^\dagger-\hat{a}_\mathrm{S_1})
    +\epsilon_\mathrm{Y1}^\mathrm{I}*\hat{\sigma}_{x\mathrm{Y1}}+\epsilon_\mathrm{Y1}^\mathrm{Q}*\hat{\sigma}_{y\mathrm{Y1}}\\
    +\chi_\mathrm{Y1S1}\hat{a}_\mathrm{S1}^{\dagger}\hat{a}_\mathrm{S1}\ket{e_\mathrm{Y1}}\bra{e_\mathrm{Y1}}
    +\frac{\chi_\mathrm{S1S1}}{2}\hat{a}_\mathrm{S1}^{\dagger}\hat{a}_\mathrm{S1}^{\dagger}\hat{a}_\mathrm{S1}\hat{a}_\mathrm{S1}.
\end{gathered}
\end{equation}

\begin{figure}[tbp]
    \centering
    \includegraphics[width=480pt]{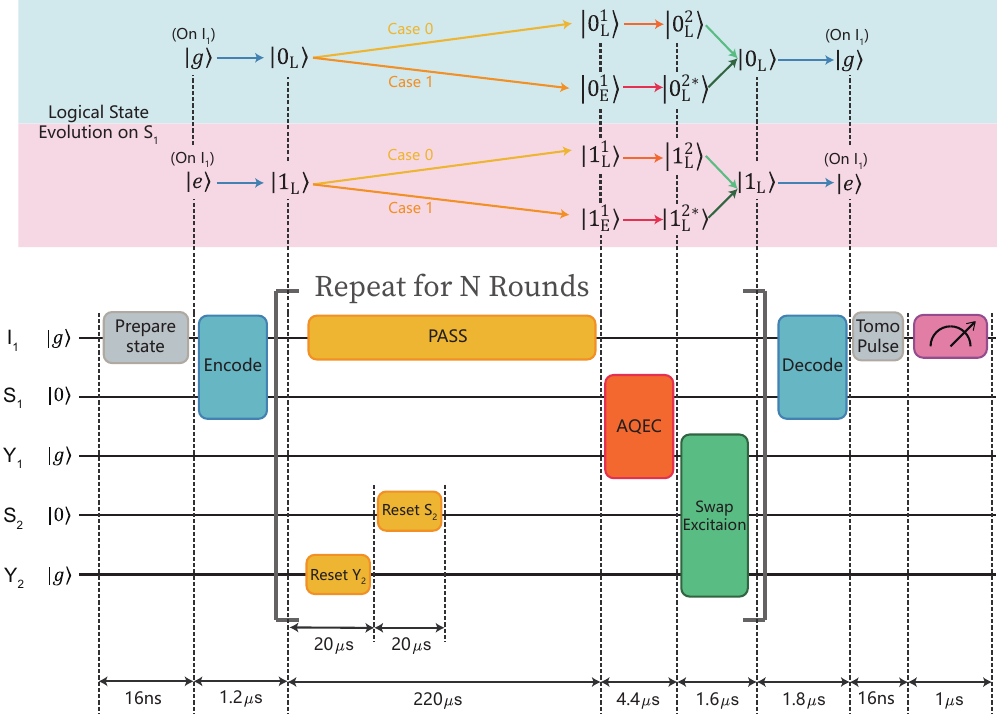}
    \caption{\textbf{Flowchart of the AQEC experiment and the corresponding evolution process of logical states.} Case 0 refers to the scenario where no single-photon loss occurs during the $220\,\mu$s free evolution, while Case 1 refers to the scenario where a single-photon loss does occur during the same period. The durations of each stage are summarized at the bottom of the figure. The top portion of the figure illustrates the evolution of the system's logical states. ``On $I_1$" indicates that the logical state is encoded in the transmon qubit $I_1$ at these times, while at other times the logical state remains encoded in $S_1$.
    }
    \label{fig:Evolution_Map_SI}
\end{figure}

We summarize the time duration of each stage in our AQEC protocol and the evolution of the corresponding logical states at each stage in Fig.~\ref{fig:Evolution_Map_SI}. The logical states $\ket{0_\mathrm{L}}$, $\ket{1_\mathrm{L}}$, $\ket{0_\mathrm{L}^1}$, $\ket{1_\mathrm{L}^1}$, $\ket{0_\mathrm{E}^1}$, $\ket{1_\mathrm{E}^1}$, $\ket{0_\mathrm{L}^2}$, $\ket{1_\mathrm{L}^2}$, $\ket{0_\mathrm{L}^{2*}}$, and $\ket{1_\mathrm{L}^{2*}}$ are defined in previous sections.

\section{Reset the qubit and the cavity with Raman process}

\subsection{Reset the qubit}
In this work, we reset $Y_2$ to its ground state by transferring its population to the short-lived readout mode $R_{Y2}$~\cite{ETHReset_Magnard_2018}. This is achieved by simultaneously driving two transitions within the joint subspace $R_{Y2} \otimes Y_2$: 1) $\ket{0,e}\leftrightarrow\ket{0,f}$ and 2) $\ket{0,f}\leftrightarrow\ket{1,g}$. Due to the extremely short lifetime of $R_{Y2}$, the transferred population rapidly dissipates, completing the reset of $Y_2$. Here, the transition $\ket{0,e}\leftrightarrow\ket{0,f}$ is achieved by directly applying a resonant drive with a frequency corresponding to $\omega_\mathrm{ef} = \omega_\mathrm{Y2}^f-\omega_\mathrm{Y2}$. For the $\ket{0,f}\leftrightarrow\ket{1,g}$ transition, we utilize a four-wave mixing process by applying a drive at the frequency specified as $\omega_\mathrm{0f1g} = \omega_\mathrm{Y2}^f-\omega_\mathrm{RY2}$ on the qubit, where $\omega_\mathrm{Y2}^f=2\omega_\mathrm{Y2}+\chi_\mathrm{Y2Y2}$ is the $f$-state frequency of qubit $Y_2$. 

\begin{figure}[tbp]
    \centering
    \includegraphics[width=450pt]{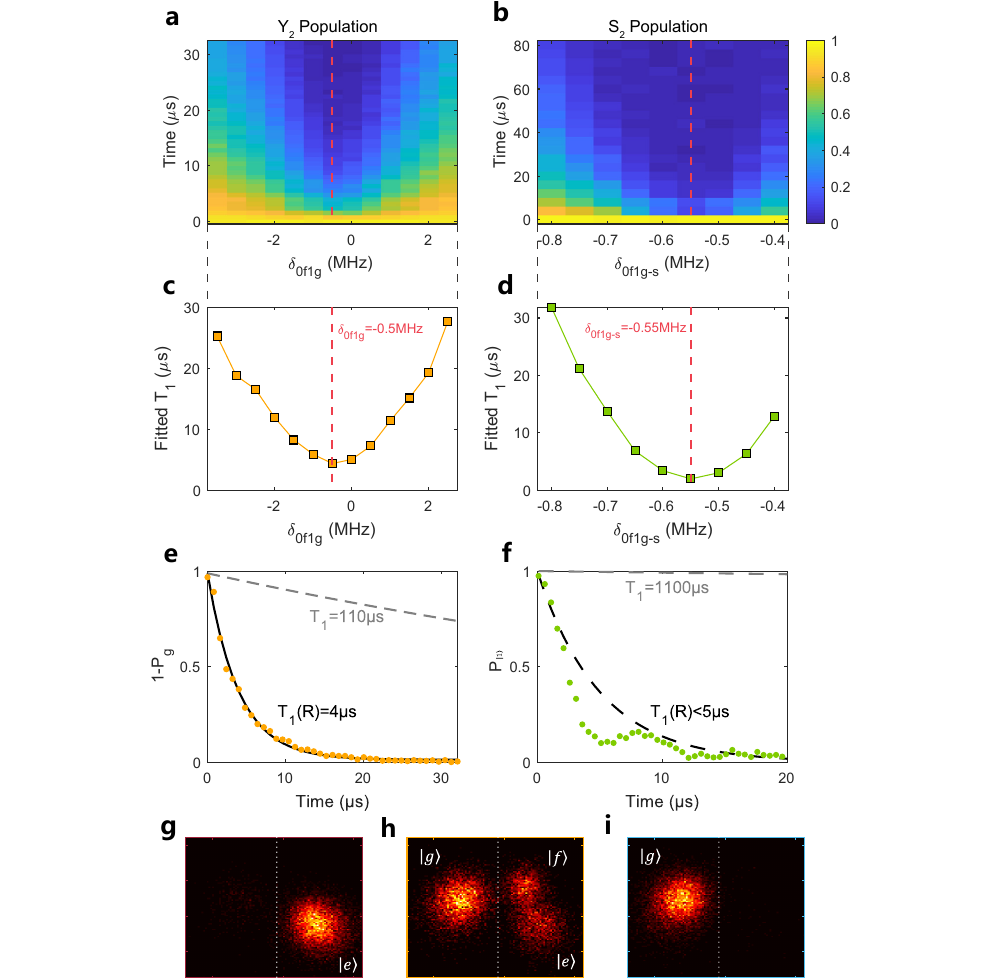}
    \caption{\textbf{Determining the reset drive frequency under the presence of a.c. Stark shifts.}
    \textbf{a.} Time-dependent evolution of the qubit population under different $\delta_\mathrm{0f1g}$.
    \textbf{b.} Time-dependent evolution of the qubit population under different $\delta_\mathrm{0f1g-s}$.
    \textbf{c.} Fitted qubit $T_1(\mathrm{R})$ under different $\delta_\mathrm{0f1g}$.
    \textbf{d.} Fitted cavity $T_1(\mathrm{R})$ under different $\delta_\mathrm{0f1g-s}$.
    \textbf{e.} Qubit population decay at the chosen working point. $T_1=110\,\mu$s refers to the intrinsic relaxation time of the mode in the absence of reset drives.
    \textbf{f.} Cavity population evolution at the chosen working point. $T_1=1170\,\mu$s refers to the intrinsic relaxation time of the mode in the absence of reset drives.
    \textbf{g-i.} 2D histogram of the readout I-Q signals at 0.1$\,\mu$s, 2.5$\,\mu$s, and 32.1$\,\mu$s of the qubit reset, respectively. The corresponding qubit state is labeled next to each bright spot in the figures. The white vertical dashed lines in the figures indicate the threshold used for the single-shot readout.
    }
    \label{fig:CalReset}
\end{figure}

To experimentally evaluate the reset efficiency, we measure the time-dependent population of the qubit states $\ket{e}$ and $\ket{f}$ while both drives are activated. Note that the bare frequencies $\omega_\mathrm{Y2}$, $\omega_\mathrm{RY2}$, and $\omega_\mathrm{Y2}^f$ are subject to the a.c. Stark shifts induced by the applied drives, rendering the values in Table~\ref{tab:allparameters} unsuitable for direct use. To compensate for this effect, we choose the drive frequency as $\omega_\mathrm{0f1g} = \omega_\mathrm{Y2}^f-\omega_\mathrm{RY2}+\delta_\mathrm{0f1g}$ in our experiments. By systematically varying $\delta_\mathrm{0f1g}$ and measuring the decay rate of the qubit population as a function of $\delta_\mathrm{0f1g}$, we obtain the results depicted in Figs.~\ref{fig:CalReset}a and \ref{fig:CalReset}c. Ultimately, we select $\delta_\mathrm{0f1g} = - 0.5\,\mathrm{MHz}$, where the qubit population decays exponentially. A fit to this exponential decay yields an estimated $T_1(\mathrm{R})=4\,\mu$s, as shown in Fig.~\ref{fig:CalReset}e. Additionally, we track the qubit state evolution through the qubit readout I-Q signals at 0.1$\,\mu$s, 2.5$\,\mu$s, and 32.1$\,\mu$s after activating the reset drive, as shown in Figs.~\ref{fig:CalReset}g-i, respectively. The data show that the qubit initially resides mostly in the $\ket{{e}}$ state, then undergoes the reset process via the $\ket{{f}}$ state, and is ultimately reset to the $\ket{{g}}$ state.

\subsection{Reset the Cavity}
Due to the imperfection of the Swap pulse, a small residual population remains in the $S_2$ cavity after the excitation transfer. Given $S_2$'s long lifetime, these residual populations persist into subsequent AQEC cycle, leading to an increased error rate in the next round of error correction. To mitigate this issue, we implement cavity reset by simultaneously applying two ``0f1g" drives. The first drive is set to the frequency $\omega_\mathrm{0f1g-s} = \omega_\mathrm{Y2}^f-\omega_\mathrm{S2}$, inducing the transition $\ket{1,{g},0}\leftrightarrow\ket{0,{f},0}$ (in the $S_2 \otimes Y_2 \otimes R_{Y2}$ subspace). The second drive operates at the frequency $\omega_\mathrm{0f1g} = \omega_\mathrm{Y2}^f-\omega_\mathrm{RY2}$, inducing the transition $\ket{0,f,0}\leftrightarrow\ket{0,g,1}$. When both drives are activated together, the population in $S_2$ is transferred to $R_{Y2}$, where it rapidly decays to the ground state, thereby achieving the cavity reset.

Taking into account the effect of a.c. Stark shifts, we choose $\omega_\mathrm{0f1g-s} = \omega_\mathrm{Y2}^f-\omega_\mathrm{S2}+\delta_\mathrm{0f1g-s}$ and systematically vary $\delta_\mathrm{0f1g-s}$ to monitor the time-dependent evolution of the Fock $\ket{1}$ population in cavity $S_2$. The results are shown in Figs.~\ref{fig:CalReset}b and \ref{fig:CalReset}d. Finally, we choose $\delta_\mathrm{0f1g-s} = -0.55\,\mathrm{MHz}$, where the cavity population is upper-bounded by an exponential decay with $T_1 = 5\,\mu s$, as illustrated in Fig.~\ref{fig:CalReset}f.

In our AQEC experiment, one round of free evolution is $220\,\mu$s. During this period, we sequentially activate a $20\,\mu$s qubit reset drive followed by a $20\,\mu$s cavity reset drive to ensure that both $S_2$ and $Y_2$ return to their ground states, as shown in Fig.~\ref{fig:Evolution_Map_SI}. We validate the effectiveness of this reset protocol using the sequence depicted in Fig.~\ref{fig:CalSwapCircuit}c. Specifically, we prepare $S_1$ and $Y_1$ in various states, apply a Swap pulse to the system, and then sequentially perform a $20\,\mu$s qubit reset, a $20\,\mu$s cavity reset, and $180\,\mu$s of free evolution. Finally, we measure the residual populations remaining in $Y_1$, $S_2$, and $Y_2$. The results are summarized in Table~\ref{tab:ResidualPopAfterFE}. The residual populations on $Y_1$, $S_2$, and $Y_2$ are all well suppressed, confirming the efficacy of our reset protocol.

\begin{table}
\begin{center}

\begin{tabular}{c|c|c|c|c|c} \hline \hline
 \textbf{Initial state on $Y_1$ \& $S_1$} & $S_1$:$\ket{0}$ & $S_1$:$\ket{1}$ & $S_1$:$\ket{2}$ & $S_1$:$\ket{3}$ & $S_1$:$\ket{4}$ \\ \hline
 \multirow{3}{*}{$Y_1$:$\ket{{g}}$} 
& $Y_1$:$0.71\%$ & $Y_1$:$0.76\%$ & $Y_1$:$0.89\%$ & $Y_1$:$0.80\%$ & $Y_1$:$0.96\%$ \\
& $S_2$:$0.72\%$ & $S_2$:$0.85\%$ & $S_2$:$0.89\%$ & $S_2$:$0.72\%$ & $S_2$:$0.97\%$ \\
& $Y_2$:$0.84\%$ & $Y_2$:$1.05\%$ & $Y_2$:$0.84\%$ & $Y_2$:$0.89\%$ & $Y_2$:$0.84\%$ \\ \hline
 \multirow{3}{*}{$Y_1$:$\ket{{e}}$} 
& $Y_1$:$0.75\%$ & $Y_1$:$0.80\%$ & $Y_1$:$0.87\%$ & $Y_1$:$0.92\%$ & $Y_1$:$0.96\%$ \\
& $S_2$:$1.06\%$ & $S_2$:$1.35\%$ & $S_2$:$1.35\%$ & $S_2$:$1.35\%$ & $S_2$:$1.61\%$ \\
& $Y_2$:$1.01\%$ & $Y_2$:$0.84\%$ & $Y_2$:$0.92\%$ & $Y_2$:$0.88\%$ & $Y_2$:$1.05\%$ \\ \hline
\hline
\end{tabular}
\caption{\textbf{Measured residual populations on $Y_1$, $S_2$, and $Y_2$ after the Swap operation and our reset protocol.} Measured population on $Y_1$, $S_2$, and $Y_2$ at the end of the sequence illustrated in Fig.~\ref{fig:CalSwapCircuit}c. The population on $S_2$ is obtained by parity measurement assisted by $Y_2$.
	\label{tab:ResidualPopAfterFE}
}
\end{center}
\end{table}

\section{AQEC without transferring ancilla excitation}


In this section, we analyze an alternative AQEC protocol where the ancilla qubit ($Y_1$) is reset directly rather than undergoing excitation transfer. To assess the impact of this direct reset approach, we simulate the reset procedure for $Y_1$ and its readout resonator $R_{Y1}$. In the simulation, we set the effective $T_1$ of $Y_1$ during the reset to 2.4$\,\mu$s, a value chosen to minimize reset-induced effects on the storage cavity and to avoid excessive thermal excitation of the qubits from overly strong drives. This $T_1$ value is derived from the effective Hamiltonian~\cite{Pechal2014PRX,Zeyt2015PRA,ETHReset_Magnard_2018} given a Rabi rate of 2.89~MHz for the $\ket{0,e }\leftrightarrow\ket{0,f}$ transition and a drive strength of 70~MHz for the $\ket{0,f}\leftrightarrow\ket{1,g}$ transition, which is set to be similar with the experimentally applied reset drive intensities on $Y_2$.

We then prepare a set of logical states in the storage cavity $S_1$ that is coupled to ancilla $Y_1$ and evaluate the process fidelity of the reset operation relative to an ideal identity process. When $Y_1$ is in the ground state, indicating no photon loss has occurred in $S_1$, the reset has no effect on the storage cavity. However, if $Y_1$ is in the excited state, indicating that a single-photon-loss error has occurred in the logical state, the reset introduces a stochastic phase error in $S_1$. This phase error arises from 1) the stochastic nature of the jump event during $Y_1$'s reset and 2) the dispersive interaction between $S_1$ and $Y_1$ that phase shifts $S_1$ whenever $Y_1$ is not in the ground state.

\begin{figure}[tbp]
    \centering
    \includegraphics[width=450pt]{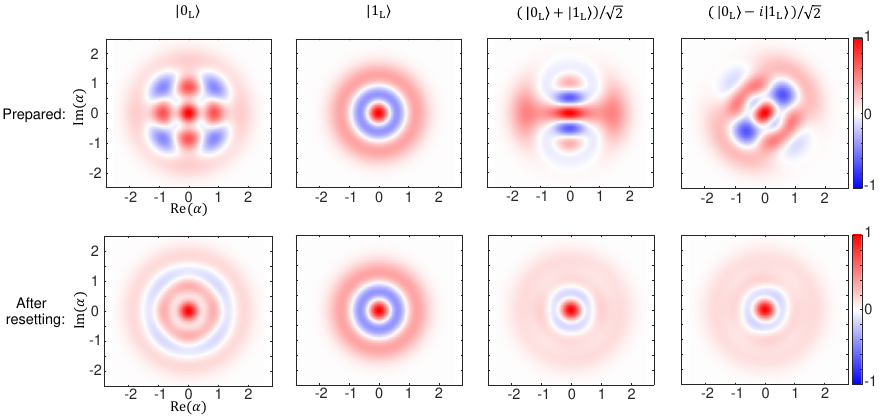}
    \caption{\textbf{Wigner functions of four different logical states of the binomial code.} The top row displays the Wigner functions of four ideal logical states immediately after state preparation in simulation. The bottom row shows the Wigner functions of the corresponding logical states after direct ancilla reset, where the ancilla is initially in the excited state in the simulation. The state fidelities of the four logical states after reset and decoding processes are 100$\%$, 100$\%$, 54.0$\%$, and 53.1$\%$, respectively. Our decoding protocol protects the fidelity of $\ket{0_\mathrm{L}}$ state against the reset-induced dephasing.}
    \label{fig:WignerAfterResetting}
\end{figure}

Our simulation results show that the process fidelity drops to 54.0$\%$ when $Y_1$ is initially excited. In this simulation, we prepare four logical states ($\ket{0_\mathrm{L}}, \ket{1_\mathrm{L} }, (\ket{0_\mathrm{L}}+\ket{1_\mathrm{L} })/\sqrt{2},$ and $(\ket{0_\mathrm{L}}-i\ket{1_\mathrm{L} })/\sqrt{2}$), each tensor-producted with $Y_1$ in $\ket{{e}}$ and $R_{Y1}$ in $\ket{0}$. We then apply the reset operation assuming a single dissipation channel of $R_{Y1}$ ($T_1 = 52\,\mathrm{ns}$; see Table~\ref{tab:allparameters}), yielding the final state of $\ket{{g}}\otimes\ket{0}$ in $Y_1 \otimes R_{Y1}$. The Wigner functions of the four storage cavity states are shown in Fig.~\ref{fig:WignerAfterResetting}. It is important to note that central symmetry in Wigner functions indicates the loss of phase information. The state fidelities of the four logical states after reset and decoding processes are 100$\%$, 100$\%$, 54.0$\%$, and 53.1$\%$, respectively.  Note that although the state $\ket{0_\mathrm{L}}$ loses the relative phase between $\ket{0}$ and $\ket{4}$ during reset, our decoding procedure still ensures that the final fidelity is not affected by reset-induced dephasing (see Eq.~\Ref{eq:DecodeWithDual}). 

For comparison, the process fidelity between a single-qubit identity operation and a total dephasing channel is 50$\%$. Similarly, for the four representative qubit states ($\ket{{g}}$,$\ket{{e}}$, ($\ket{{g}}$+$\ket{{e}}$)$/\sqrt{2}$, and ($\ket{{g}}-i\ket{{e}}$)$/\sqrt{2}$), the resulting state fidelities after total dephasing are 100$\%$, 100$\%$, 50$\%$, and 50$\%$, respectively. This shows that the reset-induced dephasing on the logical state, when $Y_1$ is excited, effectively behaves like a total dephasing channel of the form $\varepsilon(\rho)=\frac{1}{2}\rho + \frac{1}{2} Z\rho Z$.

\begin{figure}[tbp]
    \centering
    \includegraphics[width=300pt]{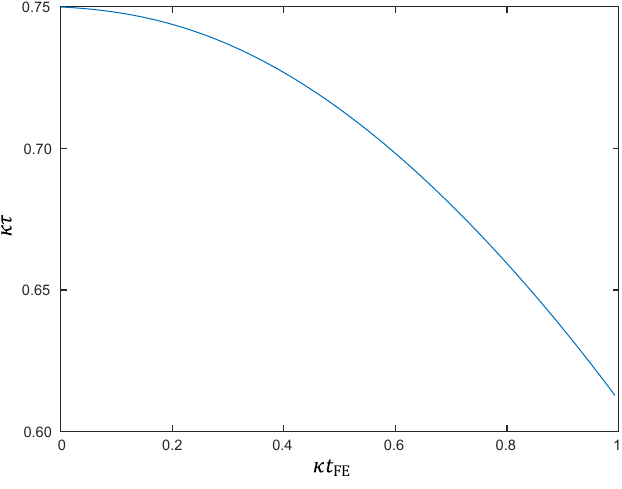}
    \caption{\textbf{Numerical results of AQEC performance with direct ancilla reset.} The vertical axis represents the ratio of the process fidelity decay time $\tau$ of the binomial code to the relaxation time $1/\kappa$ of the Fock state $\ket{1}$. The horizontal axis indicates the ratio of the waiting time $t_\mathrm{FE}$ to the relaxation time $1/\kappa$ of $\ket{1}$.}
    \label{fig:AQECwithDirectReset}
\end{figure}

Next, we examine whether the direct resetting approach enables the logical state to surpass the break-even point. Figure~\ref{fig:AQECwithDirectReset} shows numerical results of an ideal AQEC operation on a binomial code, with varying waiting times between the ideal state preparation and the ideal AQEC operation. The vertical axis represents the ratio of the process fidelity decay time $\tau$ of the binomial code to the relaxation time $1/\kappa$ of the Fock state $\ket{1}$. The horizontal axis shows the ratio of the waiting time $t_\mathrm{FE}$ to the relaxation time of $\ket{1}$. In this numerical simulation, all operations are ideal except for the reset process: if the ancilla qubit is in the excited state, it introduces a total dephasing channel to the binomial code. Thus, the probability of such a dephasing channel is equal to the probability of a single-photon loss occurring during the waiting period. 

As shown in Fig.~\ref{fig:AQECwithDirectReset}, the performance of the AQEC process degrades as the waiting time increases. This degradation results from the accumulation of uncorrectable higher-order photon loss errors over time. In the short-time limit where $\kappa t_\mathrm{FE} \ll 1$, we can estimate the process fidelity at time $t$ as: $F(t)=(1-\bar{n}\kappa t_\mathrm{FE})\times 1+\bar{n}\kappa t_\mathrm{FE}\times\frac{1}{2}$, where high-order terms of $\kappa t_\mathrm{FE}$ are neglected. The first term $(1-\bar{n}\kappa t_\mathrm{FE})\times1$ represents the contribution from trajectories with no photon loss, which are perfectly corrected by the AQEC operation, yielding a process fidelity of 1 (up to errors of order $(\kappa t_\mathrm{FE})^2$). The second term accounts for single-photon loss events, which introduce a total dephasing channel and reduce the process fidelity to 0.5. The average photon number for the binomial code is $\bar{n}=2$.

From this estimate, we can extract the process fidelity decay time $\tau$ using the approximation: $F(t_\mathrm{FE})=0.75e^{-t_\mathrm{FE}/\tau}+0.25\approx0.75(1-t_\mathrm{FE}/\tau)+0.25$. Thus, in the limit $t_\mathrm{FE}\rightarrow0$, we have $\tau=0.75/\kappa$, indicating that the optimal process fidelity decay time for the binomial code is 75$\%$ of the relaxation time of $\ket{1}$. Since the relaxation time is typically shorter than the break-even point (as shown in our experimental results in the main text), 
the binomial code cannot surpass the break-even point under the direct resetting approach, where the reset process induces almost total dephasing noise in the logical state.

As previously discussed, this dephasing effect arises from the stochastic nature of quantum jumps during the reset, combined with the dispersive coupling between the storage cavity and the qubit. Increasing the drive strength of the $\ket{0,f}\leftrightarrow\ket{1,g}$ transition can shorten the reset duration. When the ancilla qubit's effective $T_1$ is much shorter than $1/\chi_\mathrm{S1Y1}$, the dephasing effect can be suppressed or even eliminated. However, this requires extremely strong drives. Designing experimental parameters to avoid higher-order interactions and thermal excitations induced by such strong drives is crucial for determining whether AQEC with direct resetting can surpass the break-even point.

In our experiments, we observe significant performance degradation in our device when further increasing the drive strength of the $\ket{0,f}\leftrightarrow\ket{1,g}$ transition. Therefore, we abandon direct resetting in favor of a two-step protocol: we first transfer the ancilla excitation to decouple it from the logical state and then perform the reset.

\section{Error analysis}
In this section, we analyze the error budget in the AQEC experiment. We employ the normalized fidelity defined as $F=(F_\chi-0.25)/0.75$ for analysis. We categorize the errors occurring in a single AQEC cycle (comprising 220$\,\mu$s free evolution, Error Correction, and Excitation Transfer) into four distinct sources: Intrinsic Error, Recovery Error, Excitation Transfer Error, and Thermal \& Reset Error. 

The Intrinsic Error may originate from multi-photon loss, code space orthogonality degradation caused by no-jump errors, PASS-induced qubit excitation, and thermal excitation errors in the logical cavity. To evaluate this error mechanism, we simulate the coupled $I_1\otimes S_1$ system using QuTiP while considering only the dissipation of the logical cavity $S_1$. $S_1$ is first initialized in the logical states. According to the simulation results, after $220\,\mu$s of evolution with the same PASS drive applied to $I_1$ as in the experiment, $p_\mathrm{L} = 77.3\%$ of the cavity state remains in the logical subspace (denoted as case 0) with a fidelity of $F_\mathrm{FE}^\mathrm{L} = 94.4\%$, while $p_\mathrm{E} = 22.0\%$ transitions to the error subspace (denoted as case 1) with a fidelity of $F_\mathrm{FE}^\mathrm{E} = 96.1\%$. The remaining 0.7\% resides outside these two defined subspaces. 

The Recovery Error represents errors arising from imperfections in the error correction operation on the logical cavity. We estimate this error by simulating the coupled $S_1\otimes Y_1$ system using QuTiP. We initialize $S_1$ in both the logical subspace and the error subspace, respectively, and apply the same AQEC pulse as implemented in the experiment. The simulation results show that when $S_1$ is initially in the logical subspace (case 0), the recovery fidelity is $F_\mathrm{EC}^\mathrm{L} = 96.0\%$, while when $S_1$ starts in the error subspace (case 1), the recovery fidelity is $F_\mathrm{EC}^\mathrm{E} = 94.3\%$. 

The Excitation Transfer Error corresponds to errors induced by residual population from the Swap pulse. Due to the cross-Kerr coupling among $Y_1$, $S_2$, and $S_1$, the residual population remaining on $Y_1$ and $S_2$ induces dephasing in $S_1$. According to Table~\ref{tab:ResidualPop}, after excluding the effects of ground-state readout errors and qubit dephasing, the average residual population can be estimated as 0.84\% (case 0) and 2.76\% (case 1). Considering that the $\ket{0_\mathrm{L}}$ state is guaranteed to be unaffected by cavity dephasing (see Sec.~\ref{sec:AQECPulse}), the fidelity of the logical state under a full dephasing channel is 1/3. Consequently, the fidelity of the excitation transfer process can be estimated as $F_\mathrm{swap}^\mathrm{L} = 1-0.84\%*2/3 = 99.4\%$ (case 0) and $F_\mathrm{swap}^\mathrm{E} = 1-2.76\%*2/3 = 98.2\%$ (case 1). 

Thermal and Reset Errors refer to the errors induced by unwanted residual populations at the end of each free evolution cycle. These populations primarily originate from the system's intrinsic thermal excitations and the imperfection of the reset process, ultimately degrading subsequent recovery and excitation transfer operations. Specifically, $I_1$ and $S_3$ do not undergo a reset process. After a $220\,\mu$s evolution, their populations can be estimated using $n_\mathrm{I1} = n_\mathrm{th}^\mathrm{I1}(1-\exp(-t_\mathrm{FE}/T_1^\mathrm{I1})) = 0.30\%$ and $n_\mathrm{S3} = n_\mathrm{th}^\mathrm{S3}(1-\exp(-t_\mathrm{FE}/T_1^\mathrm{S3})) = 0.05\%$, respectively. Populations of $Y_1$, $S_2$, and $Y_2$ can be derived by averaging the data presented in Table~\ref{tab:ResidualPopAfterFE}, as $n_\mathrm{Y1}^\mathrm{L} = 0.55\%$, $n_\mathrm{S2}^\mathrm{L} = 0.12\%$, $n_\mathrm{Y2}^\mathrm{L} = 0.55\%$ (case 0), and $n_\mathrm{Y1}^\mathrm{E} = 0.56\%$, $n_\mathrm{S2}^\mathrm{E} = 0.60\%$, $n_\mathrm{Y2}^\mathrm{E} = 0.69\%$ (case 1), with the effects of ground-state readout errors and qubit dephasing being considered. Consequently, the total residual population of the entire system can be calculated as $n_\mathrm{res}^\mathrm{L} = n_\mathrm{I1}+n_\mathrm{S3}+n_\mathrm{Y1}^\mathrm{L}+n_\mathrm{S2}^\mathrm{L}+n_\mathrm{Y2}^\mathrm{L} = 1.6\%$ (case 0) and $n_\mathrm{res}^\mathrm{E} = n_\mathrm{I1}+n_\mathrm{S3}+n_\mathrm{Y1}^\mathrm{E}+n_\mathrm{S2}^\mathrm{E}+n_\mathrm{Y2}^\mathrm{E} = 2.2\%$ (case 1). Notably, both $I_1$ and $Y_1$ also exhibit thermal excitations during free evolution that subsequently decay back to the ground state. While these transient excitations do not disrupt subsequent AQEC operations, they induce complete dephasing of the logical cavity. The probability of this event occurring in a single round of free evolution can be calculated using:
\begin{equation}\label{eq:DephasingInfidelity}
p_\mathrm{deph} = (1-\exp(-\frac{n_\mathrm{th}^\mathrm{I1}t_\mathrm{FE}}{T_1^\mathrm{I1}}))-n_\mathrm{th}^\mathrm{I1}(1-\exp(-\frac{t_\mathrm{FE}}{T_1^\mathrm{I1}}))+(1-\exp(-\frac{n_\mathrm{th}^\mathrm{Y1}t_\mathrm{FE}}{T_1^\mathrm{Y1}}))-n_\mathrm{th}^\mathrm{Y1}(1-\exp(-\frac{t_\mathrm{FE}}{T_1^\mathrm{Y1}}))=0.7\%. 
\end{equation}

Consequently, the total infidelity caused by thermal effects is given by $N_\mathrm{th}^\mathrm{L} = n_\mathrm{res}^\mathrm{L}+2/3*p_\mathrm{deph} = 2.0\%$ (case 0) and $N_\mathrm{th}^\mathrm{E} = n_\mathrm{res}^\mathrm{E}+2/3*p_\mathrm{deph} = 2.6\%$ (case 1). Since these infidelities occur during the free evolution stage, they directly reduce the fidelity of the free evolution process. Therefore, the fidelity of a single AQEC round can be estimated as:
\begin{equation}\label{eq:TotalFidelity}
F = p_\mathrm{L}(F_\mathrm{FE}^\mathrm{L}-N_\mathrm{th}^\mathrm{L})F_\mathrm{EC}^\mathrm{L}F_\mathrm{swap}^\mathrm{L}+p_\mathrm{E}(F_\mathrm{FE}^\mathrm{E}-N_\mathrm{th}^\mathrm{E})F_\mathrm{EC}^\mathrm{E}F_\mathrm{swap}^\mathrm{E}=87.2\%. 
\end{equation}

\begin{figure}[tbp]
    \centering
    \includegraphics[width=300pt]{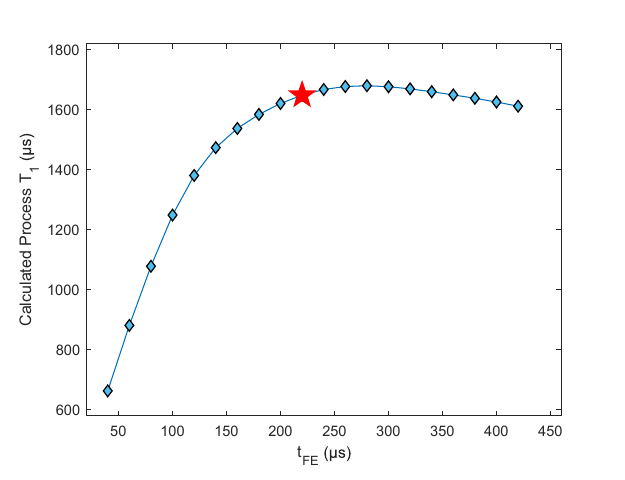}
    \caption{\textbf{Simulated process $T_1$ versus free evolution time $t_\mathrm{FE}$.} The red star represents the experimentally chosen free evolution time $t_\mathrm{FE}=220\,\mu$s.
    }
    
    \label{fig:PredictedProcT1}
\end{figure}

This result is in excellent agreement with the experimentally measured result $F_{\mathrm{exp}}=e^{-t_\mathrm{w}/T_\mathrm{proc}} = 87.0\%$, where $t_\mathrm{w}=226\,\mu$s is the time duration of a single AQEC cycle and $T_{\mathrm{proc}}=1627\,\mathrm{\mu}$s is the experimentally measured process $T_1$ time. We further simulate the process $T_1$ for the AQEC process under different free evolution times. The results are shown in Fig.~\ref{fig:PredictedProcT1}, where the experimentally chosen value of 220 $\mu$s is approximately located at the peak of the curve to achieve the optimal AQEC performance.

%